\documentclass[structabstract]{aa}
\usepackage{natbib}
\bibpunct{(}{)}{;}{a}{}{,} 
\usepackage{graphicx}
\usepackage{txfonts}
\def\kms{km s$^{-1}$}
\def\kmss{km s$^{-1}$\space}
\def\micron{$\mu$m}
\def\microns{$\mu$m\space}
\def\arcsec{$^{\prime\prime}$}
\def\arcsecs{$^{\prime\prime}$\space}

\def\arcmin{$^{\prime}$}
\def\arcmins{$^{\prime}$\space}
\def\deg{$^{\circ}$}
\def\degs{$^{\circ}$\space}

\def\n2h{N$_2$H$^+$}

\def\cii{[C\,{\sc ii}]\space}
\def\ciis{[C\,{\sc ii}]}
\def\ci{[C\,{\sc i}]\space}
\def\cis{[C\,{\sc i}]}
\def\hi{H\,{\sc i}\space}
\def\his{H\,{\sc i}}
\def\hii{H\,{\sc ii}\space}

\def\co13{$^{13}$CO\space}
\def\co18{C$^{18}$O\space}
\def\co12{$^{12}$CO\space}

\def\c+{C$^+$}

\def\h2{H$_2$}

\begin{document}

                \title{Origin and $z$--distribution of  Galactic diffuse \cii emission}

\titlerunning{Diffuse Galactic  \cii emission}
\authorrunning{Velusamy, Langer}

\author{T. Velusamy
          and
          W. D. Langer
         \fnmsep\thanks{{\it Herschel} is an ESA space observatory with science instruments provided by European-led Principal Investigator consortia and with important participation from NASA.}
          }


   \institute{Jet Propulsion Laboratory, California Institute of Technology,
              4800 Oak Grove Drive, Pasadena, CA 91109, USA\\
              \email{Thangasamy.Velusamy@jpl.nasa.gov}
             }
   \date{Received 11 June 2014 / Accepted 04 September 2014}


\abstract
{ The \cii emission is an important  probe of  star formation in the Galaxy and in external galaxies. The GOT C+ survey  and its follow up observations  of spectrally resolved 1.9 THz \cii emission using  \textit{Herschel}  HIFI provides the data needed to quantify   the  Galactic  interstellar \cii gas  components as tracers of star formation. }
{We determine the source of the diffuse \cii emission  by studying its spatial (radial  and vertical)   distributions  by separating and evaluating the fractions of \cii    and CO  emissions in the Galactic ISM gas components.      }
 {  We used the HIFI \cii   Galactic survey (GOT C+), along with ancillary \his, $^{12}$CO, $^{13}$CO,  and C$^{18}$O data toward 354  lines of sight, and several  HIFI   \cii and \ci position-velocity maps.   We quantified the  emission  in each spectral line profile by evaluating the intensities in 3 \kmss wide  velocity  bins, ``spaxels''.   Using the  detection of \cii with CO or \cis,  we separated the dense  and diffuse gas components.     We derived   2-D Galactic disk maps   using the spaxel velocities for kinematic distances.     We separated the warm and cold \h2 gases by comparing CO emissions with and without associated \ciis.   }
{We find evidence of widespread diffuse \cii emission  with a $z$--scale  distribution larger  than  that for  the total \cii or CO. The diffuse \cii emission consists of  (i) diffuse molecular (CO faint) \h2 clouds  and (ii) diffuse \hi clouds and/or WIM.     In the inner Galaxy we find a lack of \cii detections in a majority ($\sim$62\%) of \hi spaxels and show that  the diffuse component primarily comes from the WIM ($\sim$21\%) and that  the  \hi gas is not a major contributor to the diffuse component ($\sim$6\%).    The warm-\h2 radial profile   shows an excess in the range 4 to 7 kpc, consistent with enhanced star formation there.   }
{ We derive, for the first time, the 2-D \cii spatial distribution in the plane and the $z$--distributions of the individual \cii gas component.    From the GOT C+ detections  we estimate the fractional  \cii emission tracing (i) \h2 gas in dense and diffuse molecular clouds as  $\sim$48\%  and $\sim$ 14\%, respectively, (ii) in the \hi gas   $\sim$18\%, and  (iii) in the WIM $\sim$21\%.   Including non-detections from \hi increases the \cii in \hi to $\sim$ 27\%.   The $z$--scale distributions FWHM from smallest to largest are  \cii sources with CO, $\sim$130 pc, (CO-faint) diffuse \h2 gas, $\sim$200 pc, and  the diffuse \hi and WIM, $\sim$330 pc.      CO observations, when combined with \ciis,   probes  the   warm-\h2 gas,  tracing star formation.  }
{}  

\maketitle


\section{Introduction}

Ionized carbon  is widespread throughout the interstellar medium (ISM) ranging from the tenuous warm ionized medium (WIM) component  to the diffuse atomic and/or molecular hydrogen clouds, and the  photon dominated regions  (PDR) surrounding dense molecular clouds.  Thus the 1.9 THz (158 \micron) $^2$P$_{3/2}$--$^2$P$_{1/2}$ transition of C$^+$ (\ciis) is a very important tracer and diagnostic of ISM conditions. It is the strongest Galactic far-IR emission line  and, under most conditions where carbon  is ionized, is the most important coolant. The \cii emission in the ISM can be excited over a wide range of interstellar environments through  collisions  with electrons, as well as atomic and molecular hydrogen \cite[cf.][]{goldsmith2012,wiesenfeld2014}.
Thus \cii is a key tracer of the evolution of the largely atomic regions into denser, cooler, molecular clouds in which new stars are formed, and it is widely used as a tracer of star formation in the Milky Way and other galaxies \cite[e.g.,][]{Malhotra2001,contursi2002,stacey2010,braine2012,Pineda2014}. 
 The use of [CII] as a probe of galaxy evolution is   growing in importance on  all distance scales from the Milky Way to  the highest redshift galaxies   even  to the epoch of re-ionization \cite[e.g.,][]{gong2012}.

There is, however, an ongoing controversy about the origin of the bulk of the \cii emission, which has been claimed to come primarily, or collectively, from the extended low-density warm interstellar medium (ELDWIM; \cite{heiles1994,abel2006mnras}), the atomic gas \citep{Bennett1994}, and the photon-dominated region  of molecular clouds associated with massive  young stars \citep{Shibai1991,stacey2010}.   Almost all of these assertions were based on spectrally unresolved \cii surveys of the ISM, whereas only a spectrally resolved survey can locate the source of the emission with which to evaluate the relative importance of each of these ISM components throughout the Galaxy.  To date the {\it Herschel} Open Time Key Programme: Galactic Observations of Terahertz C+ (GOT C+)  is the only spectrally resolved \cii survey of the Galaxy that can be used to locate and quantify the various ISM contributions to the observed \cii intensity  \cite[cf.][]{langer2010, velusamy2010}.   The   GOT C+ data are  available in the {\it Herschel} data archives\footnote {ftp://hsa.esac.esa.int/URD\_rep/GOT$\_$Cplus/}. In this paper we use this data base to address the question of the origin of \cii emission and its distribution throughout the Galactic plane, including its {radial and} vertical  distributions and the diffuse component.    We also use the \cii and CO gas fractions to derive the radial distribution of the warm-\h2 gas, which can be regarded as a measure of the star formation rate (SFR) and/or star formation efficiency (SFE).

Earlier Galactic surveys lacked the spectral resolution to resolve the velocity features needed to study such a wide range of ISM components.  For example, the COBE-FIRAS all-sky map of the Galaxy in \cii had a 7\degs beam and velocity resolution $\sim$1000 \kmss \citep{Bennett1994},  and the BICE map, though with better  angular resolution of 15\arcmins and velocity resolution of 175 \kms, covered only the limited Galactic longitude range  350\degs $<$ $l$ $<$ 25\degs   \citep{Nakagawa1998}.  In contrast, the observations by FILM \citep{Shibai1996} covered 360\degs in longitude and latitude to $b$=$\pm$60\deg, but only surveyed a narrow strip along a great circle.  However, FILM with its $\sim$12\arcmins beam had a velocity resolution $\sim$750 \kmss that was insufficient for resolving spectral features. Thus in the post-{\it Herschel} era observations with GOT C+  and its follow-up, HIFI maps of \cii along a number of strips remain the most valuable  data set for characterizing the \cii emission in the Galaxy.  Unlike the \cii emission from bright PDRs with high density and/or UV fields,  which can be easily identified with their association with dense CO molecular clouds  (with or without star formation)  and, in some cases, with \hii regions, the \cii emission from the WIM and the diffuse  atomic and molecular hydrogen gas are more difficult to separate because there is no convenient widespread ancillary tracer to distinguish them.

 In the IRTS/FILM survey,   \cite{Shibai1996} detected a dominant component that is highly concentrated within the Galactic plane with  weaker \cii  emission slowly decreasing with Galactic latitude $b$ beyond $\sim$ 5\degs to 10\deg.  \cite{Makiuti2002} conclude that the major source of the \cii line emission observed by IRTS/FILM at high Galactic latitudes is the WIM. However, the data and the interpretation for the high latitude diffuse \cii emission seen by COBE \cite[cf.][]{Bennett1994,heiles1994} and FILM \citep{Makiuti2002} only apply  to the local ISM (within a kpc in the solar neighborhood)  and  are not at all  representative of the inner Galaxy.  Little is known about this diffuse \cii  component in the inner Galactic disk. In this paper we  use the GOT C+  survey  of \cii to separate its diffuse component using a statistical approach to identify characteristics that indicate the likeliest source of emission.

To date the GOT C+ data  has been used in   four major analyses (i) \cite{Pineda2013} used a globally integrated approach to separate \cii emission components looking at the average properties as a function of Galactocentric radius, but limited to $b$=0\degs (ii) \cite{langer2010,langer2014_II} and \cite{velusamy2010,velusamy2013} took a statistical approach to study the \cii components arising from the CO-dark \h2 gas using the \cii clouds identified in Gaussian fits to the narrow ($< $ 8 \kms) spectral line profiles  (iii) \cite{velusamy2012} analyzed weak broad \cii emission   arising from the WIM  along a spiral arm tangency, and    (iv) \cite{Pineda2014} studied the  \cii radial luminosity  as a tracer of Galactic star formation  using only the $b$=0\degs data.  In this paper we take a different approach that utilizes all the available \cii detections and non-detections in the GOT C+ data base and (i) separates out the dense and diffuse ISM contributions to \cii emission (ii) determines their spatial distribution throughout the inner Galactic plane and (iii) the distribution in $z$ of each ISM \cii component.

A common feature seen in the GOT C+  \cii velocity profiles   are a low brightness diffuse component in addition to the bright features associated with denser  molecular clouds as traced by CO.  The broad \cii line wings  are seen in the vicinity of much narrower emission features, which are  often associated with CO counterparts. Some narrow line width features are possibly associated with diffuse \h2 without CO.   However, a detailed study of this diffuse emission requires the ability to separate  the diffuse \cii emission from the dominant molecular gas component, thereby identifying  all features within the spectral line profiles.   These widespread low brightness broad \cii velocity features are indicative of large-scale Galactic diffuse \cii emission.  \cite{velusamy2012} used a subset of the data  along the tangent Scutum-Crux spiral arms tangency, where path lengths are long  and the \cii emission has broad linewings extending beyond the tangential velocities, to conclude that the widespread  \cii emission came from  the warm ionized medium (WIM).

 Our primary objective in this paper  is to understand how well \cii traces the Galactic gas components: molecular H$_2$, atomic \his,   warm neutral medium (WNM),  and ionized (WIM); and  then  quantify what fraction of the total Galactic \cii intensity traces  each of these gas components. Here  the \cii intensity fraction in \h2 refers to the fraction  of  the total \cii produced by \c+ excitation by \h2 molecular gas. For this purpose we include all \h2 gas components, that is both the dense and diffuse molecular clouds associated with and without   CO emission.  Our additional  goal is to use the results of the \cii gas components to derive their $z$--distribution. The Galactocentric radial distribution of the \cii emission  at $b$=0\degs  has been well studied using the GOT C+ data \cite[][]{Pineda2013,langer2014_II} while
less is known about its vertical  $z$--distribution.
(Note that in the rest of our paper we avoid using the term PDR for the \h2 molecular gas component of the \cii emission. Though   all molecular clouds are indeed  PDRs as in photon dominated regions or photon dissociation/destructive regions, the use of the term  ``PDR component'' is often misleading; for example  \cite{Pineda2010,Pineda2013} refers to the GOT C+ \cii in dense molecular clouds which are traced by $^{13}$CO as  the PDR component.)

 Our data consists of  GOT C+ \cii  along with the ancillary \his, $^{12}$CO, $^{13}$CO,  and C$^{18}$O  spectra toward 354   lines--of--sight  (LOS) in the inner Galaxy between $l$ = 270\degs to 57\deg.  However,  rather than using Gaussian fitted individual \cii spectral features \cite[][]{langer2014_II} or averaging azimuthally in rings \cite[][]{Pineda2013}, here we adopt a different approach for analyzing the diffuse \cii components. We calculate the intensities in 3 \kmss wide bins over the entire extent of each spectrum.  For each LOS  starting with  the \hi spectral line profile, which has the widest velocity range, we divide  the spectra into 3 \kmss wide bins and compute the \ciis, \his, and CO intensities in each bin.     Each  bin represents a unique volume in the Galaxy as specified by its $V_{lsr}$ and the LOS position, hereafter we refer to these velocity  bins as ``spaxels''.    We then make a spaxel by spaxel comparison of the  \ciis, \his, and CO emissions to identify different ISM regimes.   Because \hi is widespread and easily detected throughout the Galaxy we use the \hi spaxels to provide a measure of the  spatial--velocity "volumes" in the Galaxy sampled for the statistical analysis.  We evaluate  the individual contributions from ISM gas components to the \cii intensities using the spaxel statistics of \cii detections with respect to  CO and \hi detections and intensities. Here we also provide additional evidence of the spatial structure of the diffuse emission using HIFI \cii and \ci longitude--velocity strip maps  observed in another of our {\it Herschel}  programs.  The detection of \cii emission in the CO clouds is indicative of the presence of a warm (T $>$ 35 K) \c+ layer in them, while a non-detection of associated \cii in CO clouds indicates too low a temperature to excite \c+ to a level that can be detected by GOT C+.  Thus the warm \h2 gas is easily identified in the CO spaxel sample which traces the \h2 gas.     We show how this simple combination of the \cii and CO gas fractions in the GOT C+ data provides a useful probe of the Galactic distribution of the  warm-\h2 gas which is a measure of star formation.

We first determine correlations of the  spaxel intensities (integrated over the spaxel velocity width of 3 \kms) of \ciis, CO, and \hi and then interpret them  in terms of their association with dense molecular clouds, diffuse \h2 clouds, atomic \hi clouds, or the diffuse WIM component.  In addition to the GOT C+ data we present the results from our follow up HIFI mapping observations in \cii and   the fine--structure lines of carbon,  \cis, to further support and confirm   our conclusions regarding the contribution from the diffuse component. The spatial and velocity structure of the \cii diffuse component is brought out clearly in these $l-V$ and  $b-V$   maps obtained with HIFI cross scans (3\arcmins to 24\arcmins long) where \ci emission is used as a tracer of molecular gas.   Finally, we use the results of the  \cii gas components to derive their vertical  $z$--distribution.

Our paper is structured as follows.  The data is discussed in Section 2.  In Section 3 we construct the spatial-velocity maps, and the results of the spaxel analysis, comparing the distributions of \cii with  \hi  and CO.  In Section 4 we analyze the contributions of the ISM phases traced by \cii   and CO  gas components   with an emphasis on determining the sources of the diffuse \cii emission.   We also derive the radial profile of the warm-\h2 gas fractions which is useful as tracer of the star formation.    In Section 5 we determine the z distribution of the sources of \ciis, 
 CO clouds, diffuse molecular hydrogen clouds, and the diffuse atomic and warm ionized medium (WIM). We also calculate the \cii intensity fractions and   in  the inner Galaxy integrated in and above the plane, and use them to estimate the Galactic \cii luminosity.  We summarize our results in Section 6.


\section{\cii and ancillary data }

\begin{figure}[!thb]
\centering
\includegraphics[scale=0.35,angle=0]{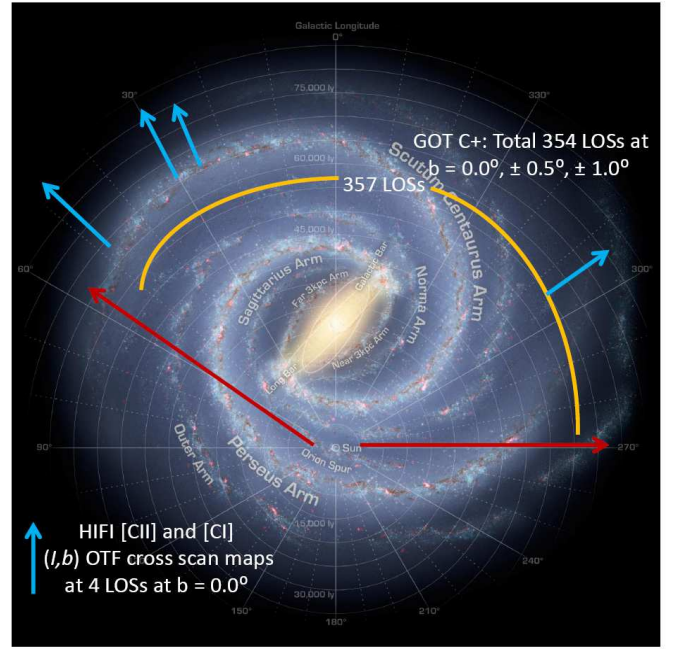}
\caption{Schematic showing the inner-Galaxy GOT C+ \cii spectral line data (from KPOT\_wlanger\_1) and the HIFI cross-scan in ($l,b$) spectral line maps of \cii and \ci  (from OT1\_tvelusam\_2). The red arrows define the longitudinal range of the GOT C+ data used here.  The blue arrows show the longitudes for the 4 cross-scan maps presented here.
}
\label{fig:fig1}
\end{figure}

The analysis in this paper uses the  \cii spectral line data from the GOT C+ survey \citep{Pineda2013,langer2014_II},  ancillary CO isotopologue and \hi observations. In addition to these we use new  \cii and \cis,
spectral line data from HIFI On-the-Fly (OTF) mapping observations presented here for the first time.  We have chosen to use only the data for the inner Galaxy, as shown by a schematic representation in  Figure~\ref{fig:fig1}, because this region of the GOT C+ survey has all the supporting observations.    The spatial and velocity resolutions for each data set are summarized in Table~\ref{tab:data_parm}.  The spatial resolution in the HIFI OTF scan maps is coarser than that for pointed observations because of the undersampling   used in this observing mode (see Section 2.2). Though the spatial resolutions for each data set vary significantly they do not affect the analysis and the results presented here.   The \ci (1-0) or CO intensities  are not used for any quantitative comparison with \cii intensities;  instead, they are only  used  for assigning  an  identification to the \cii emission of possible association with \h2 molecular gas. However, the \hi intensities are used to estimate the \cii originating in the \hi cloud layers and the effects of their beam dilution are discussed in Section 4.3.2.

\begin{table}
\caption{ Observational Data}
\label{tab:data_parm}		
\vspace{-0.25cm}
\begin{tabular}{l l c c c l}
\hline\hline
Line & Data  & beam & Velocity & ref.\\
  & Facility &size & ${\Delta}V$ &  \\
  & & & [km s$^{-1}$] & \\
\hline
[CII] & HIFI: GOT C+ & 12\arcsecs $\times$ 12\arcsec  & 1.0 & 1,2 \\
 1.9 THz &  HIFI OTF  & & & &\\
  & $l$-scan: 24\arcmins long & 80\arcsecs $\times$ 12\arcsec\ & 2.0  &1\\
    & $l$-scan: 12\arcmins long & 40\arcsecs $\times$ 12\arcsec\ & 2.0  &1\\
      & $b$-scan: 3\arcmins long & 20\arcsecs $\times$ 12\arcsec\ & 2.0  &1\\
\hline
[CI] (1-0) &  HIFI OTF  & & & &\\
 & $l$-scan: 24\arcmins long & 80\arcsecs $\times$ 46\arcsec\ & 2.0  &1\\
    & $l$-scan: 12\arcmins long & 80\arcsecs $\times$ 46\arcsec\ & 2.0  &1\\
      & $b$-scan: 3\arcmins long & 46\arcsecs $\times$ 46\arcsec\ & 2.0  &1\\
\hline
HI & SGPS/ATCA &  132\arcsecs  & 0.84 &  3\\
 &VGPS/VLA& 60\arcsecs  & 0.84 &  4\\
\hline
$^{12}$CO (1-0) & ATNF/Mopra & 33\arcsecs  & 0.8  &2\\
$^{13}$CO (1-0) & ATNF/Mopra & 35\arcsecs  & 0.8 &2  \\
$^{13}$CO (1-0) &FCRAO/GRS  & 45\arcsecs  & 0.8 &5  \\
\hline
\end{tabular}\\
$^1$This paper; \space\space
$^2$Pineda et al. (2010 and 2013);  \space\space
$^3$McClure-Griffiths et al. (2005)
$^4$Stil et al. (2006);  \space\space
$^5$Jackson et al. (2006);  \space\space
\end{table}
\subsection{GOT C+ survey data}

The \cii data used in this paper are from the GOT C+ Galactic plane [C\,{\sc ii}] survey at 1900.5369 GHz  taken with HIFI \citep[][]{degraauw2010} on {\it Herschel}  \citep[][]{pilbratt2010}. The GOT C+ \cii and the ancillary $^{12}$CO, $^{13}$CO,  and C$^{18}$O  observations and data reduction are described in \cite{Pineda2013}.  In this paper we use the \cii and CO data for 354 lines of sight toward the inner Galaxy  in the longitude range $l$ = 270\degs to 57\degs (see Figure~\ref{fig:fig1}). The corresponding H\,{\sc i} data were extracted from the VGPS survey \citep[][]{stil2006} for $l$ =14\degs to 57\degs  and SGPS \citep[][]{McClure2005} for $l$ = 270\degs to 14\deg.  The total of 354 LOS  consist of 118 longitudes with 3 LOS at each longitude alternating between $b$ = 0\deg, +0.5\deg, +1.0\degs and $b$ = 0\deg, -0.5\deg, -1.0\deg.  Note that although all the  GOT C+ \cii spectral line data used here were processed in HIPE 8, the HEB standing waves were fully corrected using standing wave shapes from off-source observations  \citep[see][]{Pineda2013}, an approach similar to that now available in HIPE 12.  Our procedure is described in more detail on the Herschel Science Centre Website under the User Provided Data Products (herschel.esac.esa.int/UserProvidedDataProducts.shtml).

\subsection{\cii \& \ci HIFI Mapping Data }

The \cii and \ci OTF scan mapping observations  were taken with the HIFI instrument toward 13 selected GOT C+ LOS in the Galactic plane at $b$=0\deg.  Here we present the results for only  4 LOS to demonstrate the presence of the diffuse \cii in the position-velocity maps. These observations were taken between October 2011 and February 2012.  At each LOS longitude  the OTF spectral line cross-scan maps were made along the Galactic longitude and latitude.    The longitude  scan ($l$-scan) lengths vary between 3\arcmins and 24\arcmin, as follows: 12\arcmins (G030.0+0.0), 24\arcmins (G045.3+0.0; G049.1+0.0; G0305.1+0.0; G345.7+0.0), and 3\arcmins for the remaining 8 LOS.  All 13 latitude scans ($b$--scans)  are 3\arcmins long.   All HIFI OTF maps were made in the LOAD-CHOP mode using a reference off-source position about 2 degrees away in latitude.  We used the off-source sky reference position from the GOT C+ program  because we have knowledge there of any \cii in the off-source spectrum.  At each longitude the OTF scans  were observed in three HIFI bands:  \ci (1--0) [$^3P_1$ -- $^3P_0$] at 492.16065 GHz, \ci (2--1) [$^3P_2$ -- $^3P_1$] at 809.34197 GHz, $^{12}$CO(7--6) at 806.6518 GHz, and  \cii [$^2$P$_{3/2}$ -- $^2$P$_{1/2}$] at 1900.5369 GHz. For a 3\arcmins OTF scan the typical observing times were $\sim$ 250s, 500s, and 2500s for \ci (1--0), \ci (2--1), and \ciis, respectively. We used the Wide Band Spectrometer (WBS) with a spectral resolution of 1.1 MHz for all the scan maps.  For the \cii OTF observations  the sampling was every 10\arcsecs  over the 3\arcmins scans,  whereas we used 20\arcsecs and 40\arcsecs samplings  for the longer 12\arcmins and 24\arcmins scans, respectively. The \ci (1--0) scans were made with half beam ($\sim$ 23\arcsec)  sampling   for all 3\arcmins scans  and 40\arcsecs sampling  for the longer scans. The reconstructed images shown in Figs. 2 \& 3 were restored with effective beam sizes corresponding to  twice the sampling length along the scan direction, which accounts for the undersampling used for these OTF scans \cite[][]{mangum2007}.

The \cii spectral line data  were taken with HIFI Band 7 which utilized Hot Electron Bolometer (HEB) detectors.  These HEBs produced strong electrical standing waves with characteristic periods of $\sim$320 MHz, that depend on the signal power.  The HIPE Level 2 \cii spectra  show these residual waves.   We found that applying the {\it fitHifiFringe}\footnote{http://herschel.esac.esa.int/hcss-doc-12.0/index.jsp\#hifi\_um:hifi\-um section 10.3.2}  task to the Level 2 data produced  satisfactory baselines. However, removal of the HEB standing waves has remained a challenge up until the recent release of HIPE-12, which  includes a new tool {\it HebCorrection}\footnote{http://herschel.esac.esa.int/hcss-doc-12.0/index.jsp\#hifi\_um:hifi\-um section 10.4.5} to remove the standing waves in the raw spectral data by  matching the standing wave patterns (appropriate to the power level) in each integration using   a database of spectra at different power levels (see Herschel Science Center (HSC) HIPE-12 release document for details). We used this HSC script to apply {\it HebCorrection} to recreate the final pipeline mapping products presented here. Any residual  HEB and optical standing waves in the reprocessed Level 2 data  were minimized further by  applying  {\it FringeFit} to the ``gridded'' spectral data (we took additional precaution in {\it FringeFit} by disabling  {\it DoAverage}  in order not to bias the spectral line ``window'').  The H-- and V--polarization data were processed separately and were combined only after applying {\it FringeFit} to the gridded data. This approach minimized the standing wave residues in the scan maps, as the standing wave  differences between H-- and V--polarization were fully  taken into account. We found that in the \cii maps produced with and without  {\it HebCorrection} all the main features, including the diffuse low brightness emissions, were nearly identical.  However, as expected,  the noise level and baselines  were better when we applied {\it HebCorrection}.  Next we generated the  $l$--$V$  and  $b$--$V$  maps from the gridded spectral  line data cubes, all created in HIPE 11 and HIPE 12 for \ci and \ciis, respectively. The matching   $l$--$V$  and  $b$--$V$ maps in \hi were made for  all LOS using the  VGPS data.  For a comparison of CO with the \ci data  we also produced  $l$--$V$  and  $b$--$V$ maps  in $^{13}$CO(1-0)  for one LOS  using the Galactic Ring Survey (GRS) \citep{Jackson2006}.


\section{Results}

\subsection{Position-velocity maps and spectral line-wings: Evidence for diffuse \cii emission}

\begin{figure}[!ht]
\hspace{-0.75cm}
\includegraphics[scale=0.5]{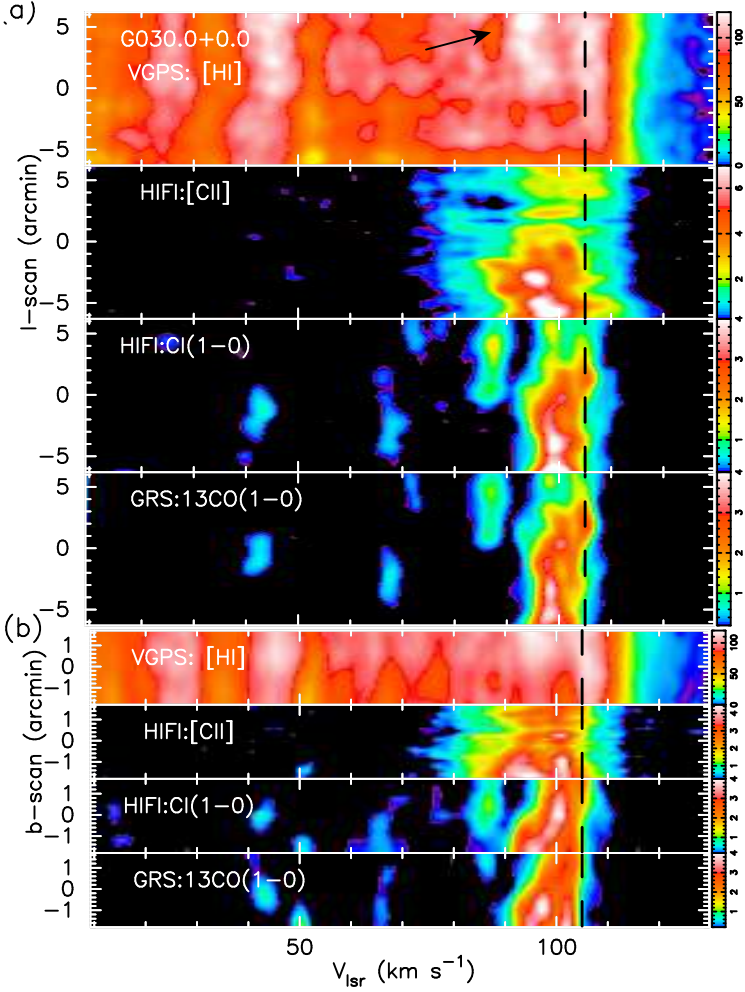}
\caption{Examples of $l$--$V$ and $b$--$V$ maps in \his, \ciis, \cis, and $^{13}$CO emission for LOS G030.0+0.0: (a) $l$--$V$ maps; and, (b)  $b$--$V$ maps. The $l$--scans are at $b$=0\deg; $b$--scans are at $l$=30.0\deg.  The intensities are  in main beam antenna temperature  (T$_{\rm mb}$)   with values  indicated by the color wedges.  These show low brightness diffuse  \cii emission along with the brighter narrow velocity features.   Comparison of the spatial and velocity structures in the \cii maps with those of \ci and  $^{13}$CO delineates the \cii arising from the denser \h2 molecular gas from the diffuse component. The vertical dashed line represents the tangential velocity for these regions. The velocity resolution in all maps is 2 \kms.  The spatial resolutions are given in Table~\ref{tab:data_parm}.   The black arrow in the top panel  marks an \hi self-absorption (HISA) feature. }
\label{fig:lbvmaps1}
\end{figure}

In this Section we present examples of position-velocity maps of several gas tracers that demonstrate the prevalence of a diffuse \cii component.    While \cii emission arises from several gas environments, its  association, or lack of association, with \ci and CO identifies the likely ISM phase from which it arises. In Figures~\ref{fig:lbvmaps1}  we show an example of the   longitude (12\arcmin) and latitude (3\arcmin) scan maps for $l$=30.0\degs and $b$=0.0\degs (labeled G030.0+0.0 in the figure).  Though similar HIFI observations exist for several LOS, here we limit our study of these maps in order  to highlight the position-velocity structure of the \cii diffuse component and that associated with \ci and CO.  Detailed discussions  of the individual features in these and other map data  will be presented elsewhere.  It can be seen in Figure~\ref{fig:lbvmaps1} that there are differences in the spatial and velocity structure among the \ciis, \cis, CO, and \hi emissions.  The spectral line intensities are shown in color scale (with color stretches  indicated by the wedges) with longitude along the Y-axis and velocity ($V_{lsr}$) along the X-axis. The strongest  emissions are seen   near the tangential velocities (marked by the dashed vertical line) corresponding to the longest path lengths through the spiral arm.

\begin{figure}[!ht]
\centering
\includegraphics[scale=0.43 ,angle=-90]{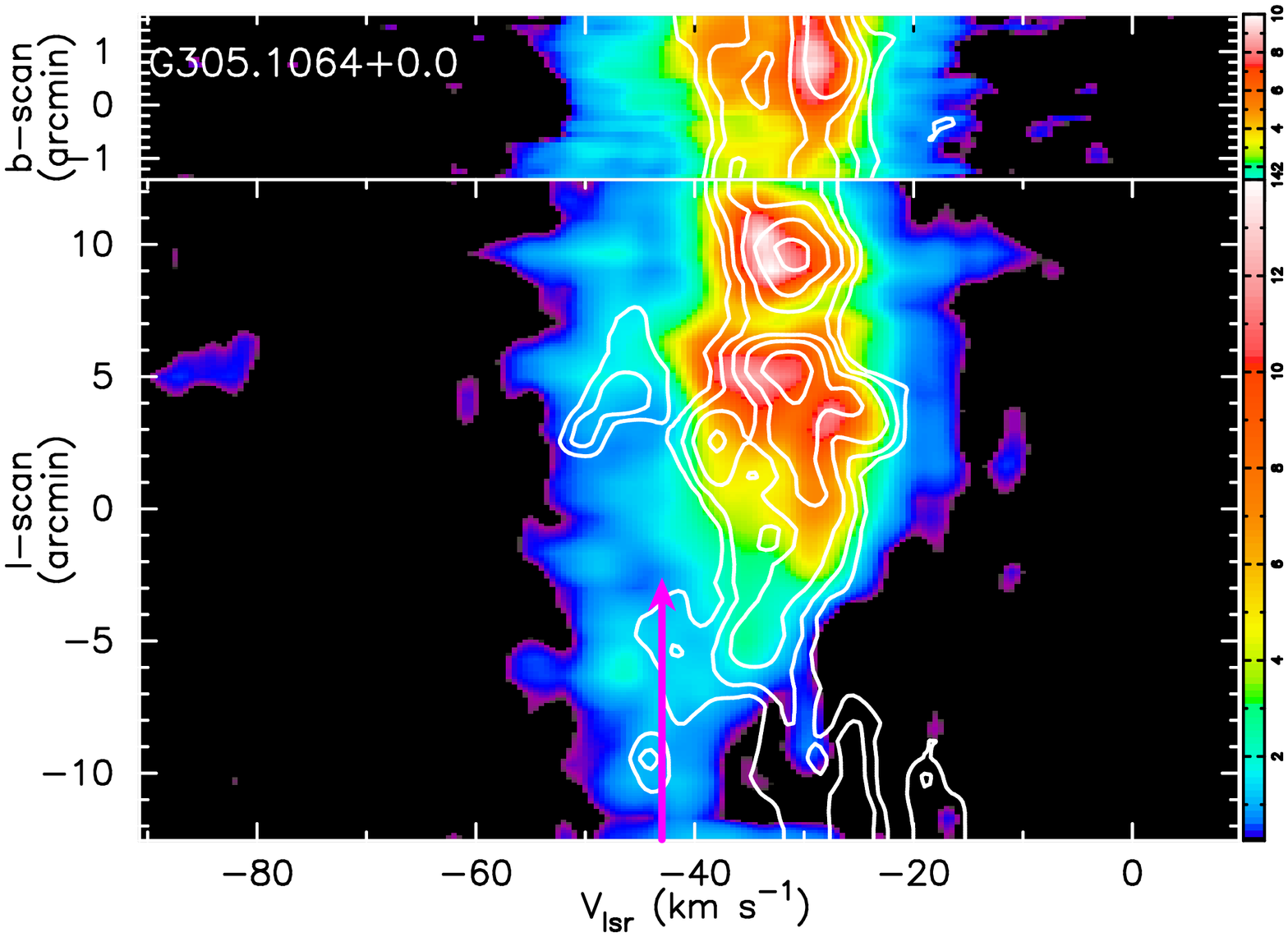}
\includegraphics[scale=0.43 ,angle=-90]{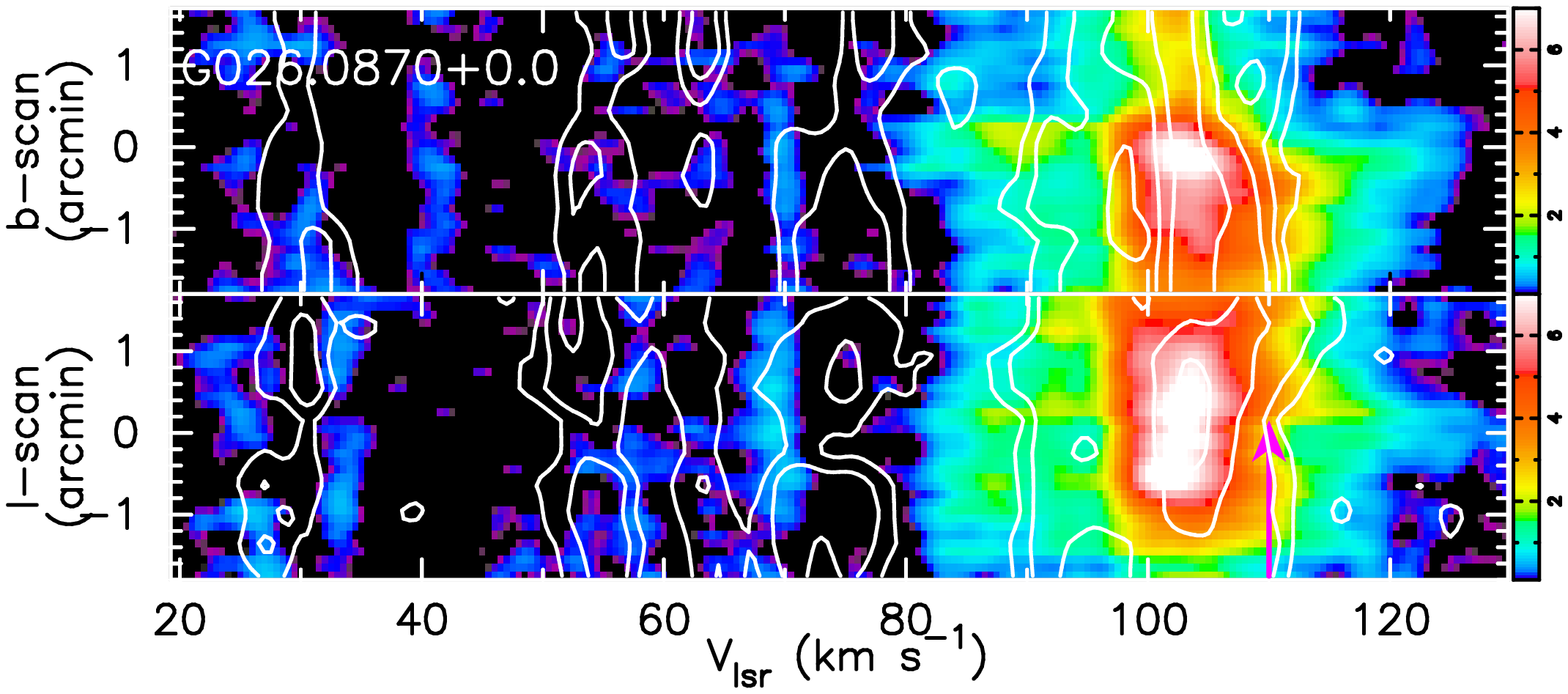}
\includegraphics[scale=0.43 ,angle=-90]{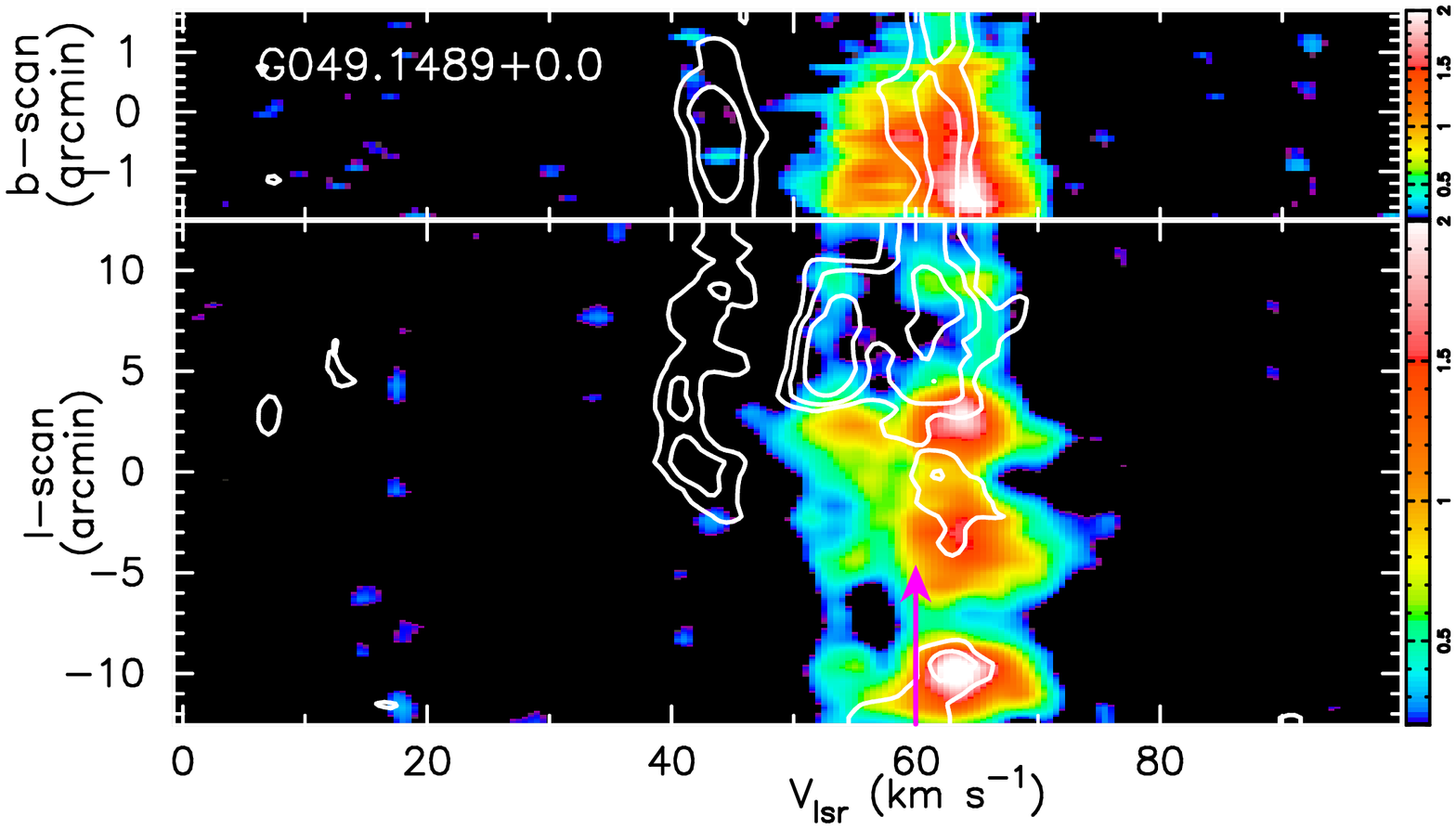}
\caption{Examples of ($l$--$V$) and ($b$--$V$) cross scan  maps of \cii and \ci along 3 LOS: G305.1+0.0 (upper panels), G026.1+0.0 (middle panels), and G049.1+0.0 (lower panels).   The $l$--scans are at $b$=0\deg; $b$--scans are at corresponding longitudes given in the images.   The \cii emission (in units of main beam antenna temperature,  T$_{\rm mb}$)   are displayed in color  with values  indicated by the color wedges.  The \ci emission (T$_{\rm mb}$)  contours    are overlaid on the  \cii  maps, and their contour values are 0.2, 0.4, 0.8, 1.6 and 3.2 K. These maps show the low brightness diffuse  \cii emission along with the brighter narrow velocity features.   The maps are centered at the indicated longitudes in each panel.    The \ci contours identify the underlying  \cii emission as  those associated with the  denser \h2 molecular gas, while \cii emission without associated \ci indicates a diffuse gas component.  The vertical arrows mark the tangential velocities for each region.}
\label{fig:lbvmaps2}
\end{figure}

As seen in Figure~\ref{fig:lbvmaps1}, there is an excellent correspondence between the HIFI  \cis(1--0) and the  $^{13}$CO(1--0) emission.
   This correspondence demonstrates that both lines trace the   presence of  \h2 gas equally well. Comparison of the spatial and velocity distributions of the \cii emission with those of \ci and $^{13}$CO, however, clearly differentiates two sources of the \cii emission, namely, (i)   the \h2 gas layers  around the dense shielded components  traced by \ci and $^{13}$CO  and  (ii)  a diffuse gas component which is not traced by \ci or $^{13}$CO.  Therefore, we can use the HIFI \cis(1--0) data   as a marker for \h2 gas, and \ci has added importance as an indicator of whether there is  \h2 present in the diffuse transition clouds in which there is  no detectable CO.  In Figure~\ref{fig:lbvmaps2} we show more examples of the $l$-- and $b$--scan maps at three  other LOS.    For   clarity of the  display in Figure~\ref{fig:lbvmaps2} we only show   HIFI \cii and \cis(1--0) maps. In each panel the \cii emission is shown as a color image while the \ci is  shown as contours  overlaid on the \cii image.

As can be seen in Figure~\ref{fig:lbvmaps1}, there is little correlation of the \cii or \ci with the large scale distribution of \his. In all our  $l$--$V$ and $b$--$V$  maps toward 13 LOS (not shown)   \hi is widespread over a broad velocity range, whereas \cii and \ci  are confined to a much narrower velocity range.   An \hi self-absorption (HISA) feature seen in this map is marked by the arrow in the top panel in Figure~\ref{fig:lbvmaps1}.  This foreground \hi absorption  is clearly detected in  emission in  both \ci and $^{13}$CO, but there appears to be a weak absorption feature in \ciis.  The presence of absorption in the foreground cloud indicates that the excitation temperature of \c+ is low, implying a low T$_{\rm kin}$ and/or density. In Figures~\ref{fig:lbvmaps1} and \ref{fig:lbvmaps2}  all emission features show enhancement near the tangent velocities,   largely  due to the longer path lengths \cite[][]{velusamy2012}.  At velocities beyond the tangent velocity no CO or \ci is observed but both \hi and \cii show significant emission.

The images and contours in Figures~\ref{fig:lbvmaps1} and \ref{fig:lbvmaps2}  bring out the similarities and differences between \cii and \ci (as a proxy for CO) spatial and velocity structures.    The  $l$--$V$ and $b$--$V$ maps  in Figures~\ref{fig:lbvmaps1} and \ref{fig:lbvmaps2}, as well as those for the other LOS,  show the following characteristics of \cii in relation to the spatial and velocity structure of the  \ci emission:
\begin{itemize}
\item \hi has emission over the widest velocity range with only a small velocity range overlapping \cii and/or \cis;
\item \ci emission has a relatively narrower velocity range than \ciis;
\item \cii shows bright narrow velocity features roughly coincident with those in \cis;
\item the broad angular and velocity extent of diffuse low surface brightness \cii components are prominent in all maps;
\item several \ci features without \cii counterparts are also present in the maps and these probably represent extremely low  excitation conditions in the molecular clouds  in which \c+, although present, is insufficiently excited to radiate  (at the sensitivity of HIFI maps).
\end{itemize}
\noindent  As can be seen in these figures there is a significant fraction of \cii emission in the diffuse component    and  this result raises the question of what is the nature of the diffuse \cii component of the ISM.  Does it come from the cold neutral medium (CNM), the warm neutral medium (WNM), or the warm ionized medium (WIM), or some combination of these sources?   We examine this question in Section 4.3 by estimating in the GOT C+ data how much \cii emission is in (a) diffuse (CO faint) \h2 gas, (b) \hi gas, and (c) the WIM.

\begin{figure}[!ht]
\centering
\includegraphics[scale=0.475,angle=0]{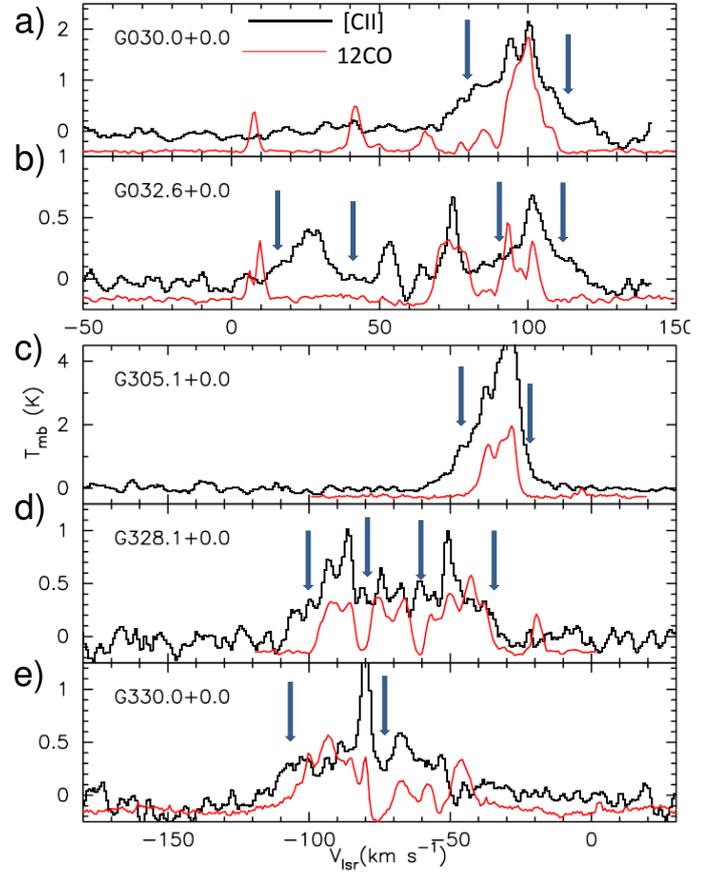}
\caption{Shown here are examples of broad line wings observed in GOT C+ spectra indicating the presence of a  diffuse \cii component.  To help identify the \cii associated with \h2 gas the $^{12}$CO spectral line profiles   are also shown. Only the \cii intensities are labeled and CO intensities are scaled down arbitrarily to fit under the \cii profiles.  The line wings  (diffuse \cii components) are indicated by arrows. See also the corresponding $l$--$V$ maps in Figures~\ref{fig:lbvmaps1} and \ref{fig:lbvmaps2}  for panels (a) and (c)  which show the spatial extent of the line-wings (diffuse components).}
\label{fig:linewings}
\end{figure}

   The examples of the spatial--velocity  maps in Figures~\ref{fig:lbvmaps1} and \ref{fig:lbvmaps2} show the prevalence of the diffuse \cii  emission which can be seen (i) in position--velocity maps ($l$-$V$ \& $b$-$V$ maps) as extended both spatially and in velocity and (ii) as broad wings in single spectral line profiles in the GOT C+ survey (see Figure~\ref{fig:linewings}). In this paper we   use  the spatial maps   only for demonstration purposes to highlight the presence of the diffuse \cii component.  For all quantitative analysis we use the spectral line profiles in the GOT C+ data.
The evidence for a significant fraction of \cii arising in a diffuse component, as seen in the spatial--velocity maps, is also abundantly clear in the intensities of the \cii line-wings. In Figure~\ref{fig:linewings}  we show  examples of \cii line profiles at five LOS in the inner Galaxy.  In Figure~\ref{fig:linewings} we also show the  $^{12}$CO line profiles to help delineate the broader line wings of \ciis.  The line wings marked in panels (a) and (c) in Figure~\ref{fig:linewings} can be traced as extended diffuse emission in the   $l$--$V$  maps shown in Figures~\ref{fig:lbvmaps1} and \ref{fig:lbvmaps2}. Thus the large (354 LOS) sample  of GOT C+ spectra offer an opportunity to examine in great detail the diffuse component both in and out of the plane (at $b$=0.0\deg, $b$ =$\pm$0.5\deg, and $b$  =$\pm$1.0\deg),  and  provides much better statistics to compare the \cii diffuse component   with \hi than the limited OTF map data.  A detailed comparison of \hi with both \cii and CO   helps us resolve the  question of the fraction of \cii originating from \h2 gas, diffuse \his, and the WIM.

\begin{figure}[!ht]
\centering
\includegraphics[scale=0.32,angle=0]{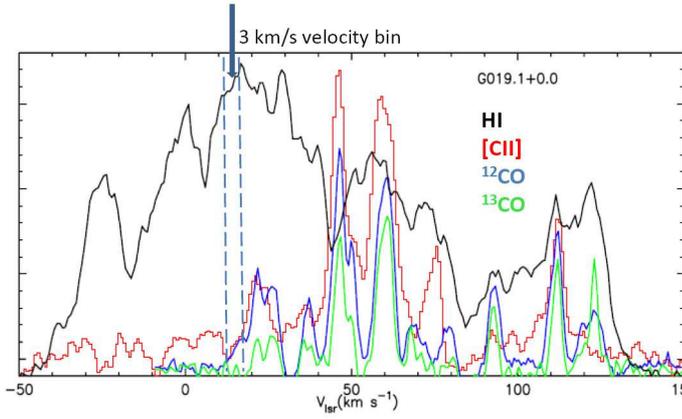}
\caption{An example of the GOT C+ spectral line data  and spaxel analysis of \ciis, \his, and CO spectra toward $l$ = 19.1304\degs and $b$ = 0.0\deg.  Each  ``spaxel''   corresponds to  a 3--\kmss wide velocity bin, as indicated schematically by the downward arrow  and delineated by the dashed lines. The intensity scales  for the spectra  are not shown.}
\label{fig:data}
\end{figure}

															
\begin{table*}[htbp]
\caption{Summary of the spaxel analysis of spectral line profiles}
\label{tab:spaxel_analysis}																
		\begin{tabular}{lcccccccccc}																	
\hline	
Item & Total & \multicolumn{3}{c}{\his$^\dag$ Detections with}  & \multicolumn{2}{c}{\cii Detections} &	\multicolumn{2}{c}{$^{12}$CO Detections} &  $^{13}$CO \\															
 & \hi &   \cii &   $^{12}$CO  &   \cii and CO & total & with CO & $^{12}$CO &with \cii & \\			
\hline
3--$\sigma$ detection limit K \kmss & 7.0 & 7.0 &7.0 & 7.0 & 0.5 & 0.5 &  3.0 & 3.0 & 0.5 \\
\hline
All data (354 LOS) & &  &  & & & & &  & \\
Total No. of spaxels  &   23229 &    6484 &    4421 &    2705 &    6484 &    2705 &    4421 &    2705 &    3653\\
Intensity (K \kmss) &  3336 &  1501 &  1117 &   794 & 10317 &  5811 & 61549 & 43236 &  8159\\
\hline
Spaxels with R$_G$ $<$ 8.5 kpc & & & & & & & & & \\
Total No. of spaxels &   15278 &    5711 &    4197 &    2609 &    5711 &    2609 &    4197 &    2609 &    3448\\
at $b$ = 0.0\degs   &    5198 &    2356 &    2096 &    1388 &    2356 &    1388 &    2096 &    1388 &    1792\\
at $b$ = $\pm$0.5\degs&    5363 &    1948 &    1340 &     826 &    1948 &     826 &    1340 &     826 &    1078\\
at $b$ = $\pm$1.0\degs &    4717 &    1407 &     761 &     395 &    1407 &     395 &     761 &     395 &     578\\
Intensity (K \kms) &    &  &  &  & & & & & \\
Total &	  2524 &  1377 &  1059 &   765 &  9132 &  5501 & 58790 & 41979 &  7807\\
at $b$ = 0.0\degs  &   985 &   637 &   548 &   422 &  4272 &  3006 & 31913 & 23170 &  4361\\
at $b$ = $\pm$0.5\degs &   883 &   465 &   338 &   239 &  2966 &  1692 & 18082 & 13273 &  2443\\
at $b$ = $\pm$1.0\degs  &   656 &   275 &   173 &   103 &  1894 &   803 &  8795 &  5536 &  1004\\
Spaxels without  distance ambiguity &    &  &  &  & & & & & \\
(used for $z$--distribution) &    &  &  &  & & & & & \\
Total No. of spaxels  &   11738 &    4484 &    3380 &    2162 &    4484 &    2162 &    3380 &    2162 &    2854\\
Intensity (K \kms)  &  1918 &  1098 &   868 &   644 &  7425 &  4629 & 49369 & 35773 &  6722\\
\hline		
\end{tabular}
\\	
$^\dag$All \hi intensities are in units of 10$^3$ K \kms.													
\end{table*}

\subsection{Statistical approach of spaxel analysis}

In this paper we analyze the statistical properties of an ensemble of spaxels to separate the contributions from the different \cii components.    To illustrate this approach we consider in detail  one of the LOS \cii spectrum along with all the ancillary \hi and CO spectra, as shown  in Figure~\ref{fig:data}.  One can see that \hi covers the entire velocity range shown, while $^{12}$CO  and $^{13}$CO are more narrowly confined and \cii covers a broader range than CO, but less than \his.  To generate the spaxel data for this LOS we calculate the intensities in 3--\kmss wide bins  and  identify those bins that have a minimum 3-$\sigma$ detection.  Along this LOS we now have $\sim$65   of the  3--\kmss spaxels, which we sort  by whether they contain various combinations of gas tracers.  We   use these   combinations  to quantify the nature of the detections in each spectral line.

The results of the spaxel analysis of all LOS are summarized in Table~\ref{tab:spaxel_analysis}, where we list the number of spaxels containing detections of the different gas tracers for all LOS.  Note that a majority of the spaxels with the \hi detections do not have \cii and/or CO detections.  As discussed in \cite{Pineda2013}, while \hi emission extends much further out in the Galaxy, \cii and CO emissions are confined to the inner Galaxy within Galactocentric radius R$_G$ $<$ 10 kpc.  Therefore we expect that  a large number of spaxels with \his, but no \cii or CO,
(see Figures~\ref{fig:distance} and ~\ref{fig:2-D_map}  below), are likely to be located at R$_G$ $>$ R$_{\odot}$ (=8.5 kpc) and  to keep our sample  homogeneous  they should be excluded for all  statistical interpretation of the \hi and \cii intensities.

\begin{figure*}[!ht]
\centering
\includegraphics[scale=0.45,angle= 0]{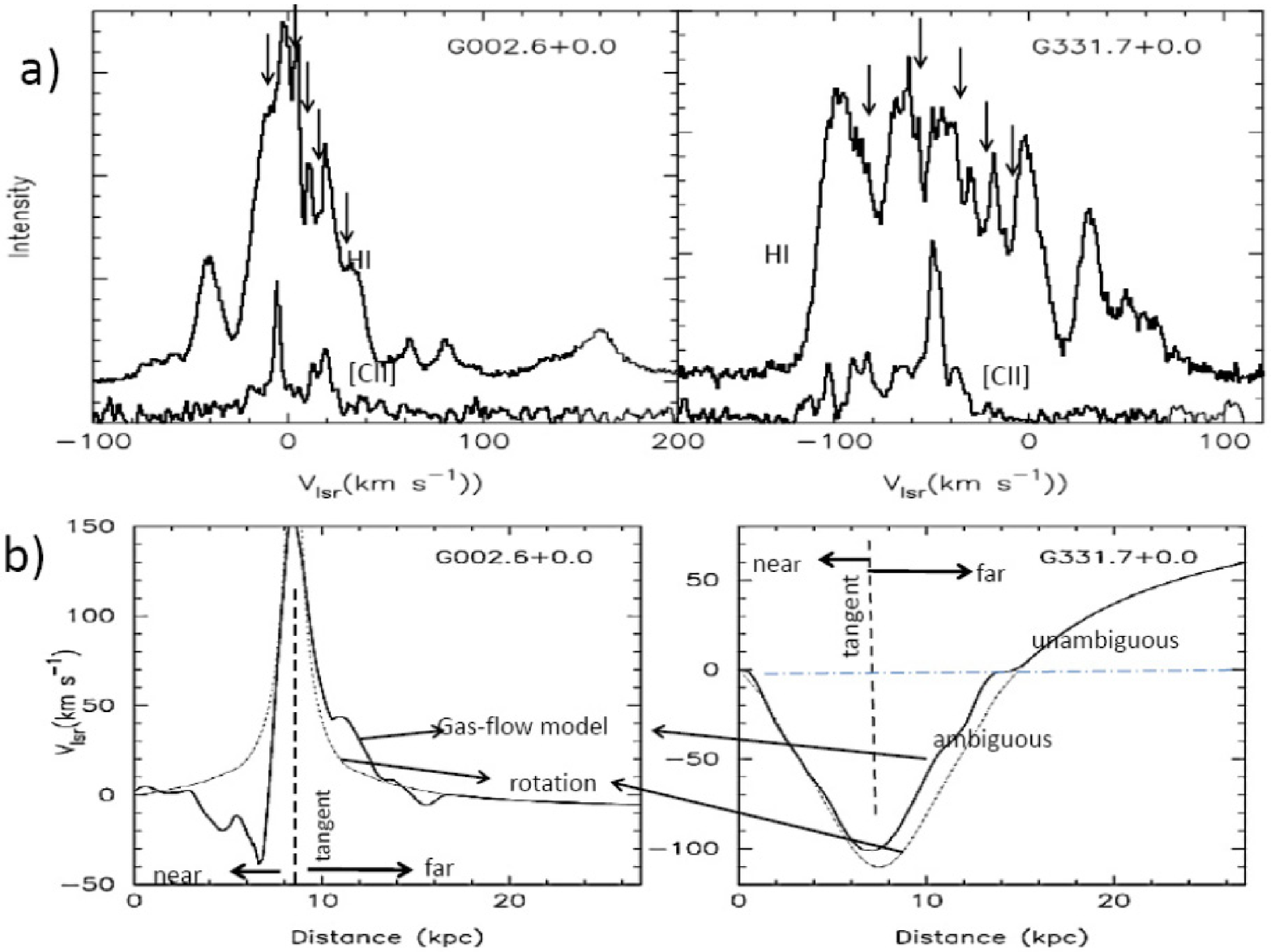}
\caption{(a) Examples of \cii and \hi spectra at two longitudes.  The downward arrows over the \hi velocity profiles represent the \hi absorption features used to resolve the distance ambiguity (see text).  The temperature scales for \hi and CO are not indicated.
(b) The $V_{lsr}$ distance relation is shown for the longitudes of the spectra in panel (a).  The dotted line  represents the solution for a standard Galactic rotation model. The solid line represents that for standard Galactic rotation supplemented by  the gas-flow model (see text).  The vertical dashed line indicates the tangent distance.
\label{fig:distance} }

\centering
\includegraphics[scale=0.55,angle=-90]{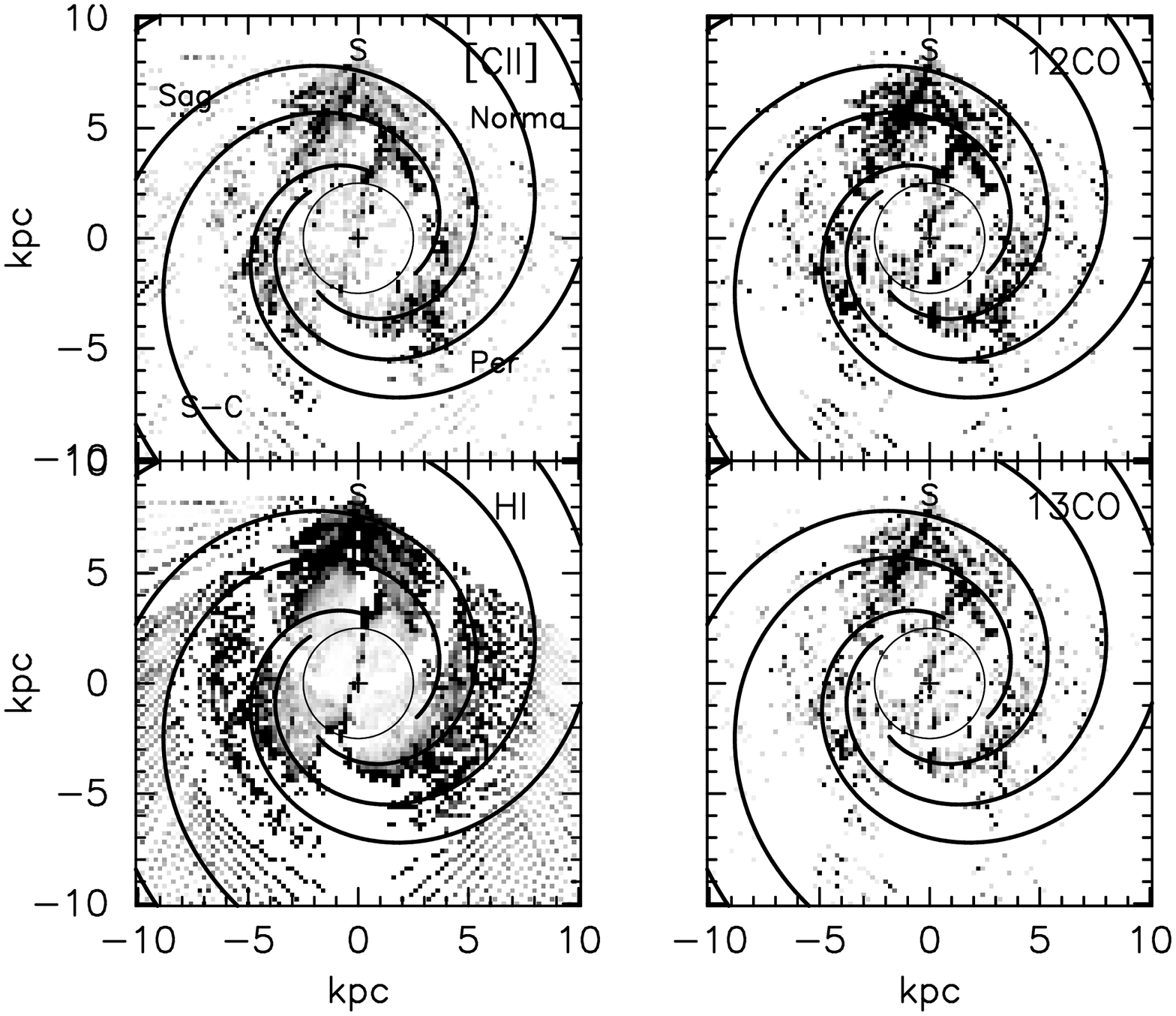}
\caption{2-D maps of Galactic \cii emission compared with that  of \hi and CO  emissions in the spectral line data in the GOT C+ survey. Here we use all  spaxels with    distance estimates (see text). We use only intensities in the spaxels  with \his, \ciis, or  CO detections. The spiral arms are identified in the top left panel and S denotes the solar  system. The circle (2 kpc radius) marks the Galactic center region.   The void seen near the tangent points in all maps is an artifact resulting from the algorithm used to resolve the near--  \space far--distance ambiguity (see text). }
\label{fig:2-D_map}
\end{figure*}

\subsubsection{Spaxel distances}
\label{sec:spaxel_distance}

The velocity resolved spectral data provide kinematic distances to each spaxel.  We can also use this information to derive the $z$--distribution  of the ISM gas components, which is important for determining the Galactic \cii luminosity \cite[see][]{Pineda2014}. We derive the distances to each spaxel from its $V_{lsr}$  using   a rotation curve based on the hydrodynamical models of \cite{pohl2008}  following the procedure discussed by  \cite{Johanson2009} and using the code given to us by C. Kerton (private communication).  This approach provides a more accurate kinematics of the clouds near the Galactic center and thus more realistic distances for the longitudes $|l|$ $<$ 6\degs than that using a simple rotation curve.  In Figure~\ref{fig:distance} we show examples of  \cii and \hi spectra for two LOS ($l$ = 2.6\degs and 331.7\deg)   showing the spaxel range  $V_{lsr}$  along with the  derived  $V_{lsr}$ distance plots for Galactic rotation  with and without a gas-flow model component. We assumed the  rotation curve given by \cite{Johanson2009}, the distance of the Sun from the Galactic center, R$_{\odot}$ = 8.5 kpc, and an orbital velocity of the sun with respect to the Galactic center, $V_{\odot}$ = 220 \kms.  Thus for longitudes closer to the Galactic center the gas-flow model allows us to use all the observed velocity components, including those which are ''forbidden`` in a simple rotation curve. However, for $|l|$ $>$ 6\degs there is little difference between the simple rotation and gas-flow models.  We use the ($V_{lsr}$--Distance) plots obtained for  each LOS longitude and then interpolate the distances for each spaxel $V_{lsr}$  in the spectra.  As indicated in the bottom panel of Figure~\ref{fig:distance} in all the LOS spectra, for a range  of $V_{lsr}$ corresponding to distances within the solar circle, the  Galactic  rotation models yield two solutions, a near-- and a far--distance.  However, each $V_{lsr}$ represents a unique value for the Galactic radial distance (R$_G$).  Therefore for any analysis that requires only R$_G$ we can use all the spaxels, and    the number of spaxels detected within R$_G$ $<$ 8.5 kpc are summarized in Table~\ref{tab:spaxel_analysis}. For any other  analysis that requires  distances we   use only the subset of   spaxels for which we have unambiguous, or relatively high confidence, in their  values and those for which the  near--  \space far--distance ambiguities can be  resolved.  Near the tangencies where the near-- and far--distances are within 2 kpc   we can use their average distances, thus these spaxel distances are good to within $\pm$1 kpc.

  Of the total of 15278 spaxels within R$_G$ $<$ 8.5 kpc we have unambiguous distances for 3537 spaxels ($\sim$23\%)  located within 1 kpc of the tangency and another 1108 spaxels have unambiguous far--distance solutions. For the  remaining  10633 spaxels,
wherever possible, we  use the \hi self absorption (HISA) features  in the \hi spectra to resolve the distance ambiguity by following the procedure of \cite{Duval2009}.  Wherever \hi absorption is detected at the spaxel $V_{lsr}$ we adopt the near distance solution.  The distances for  spaxels without HISA detections are uncertain, and therefore, they are not used in our analysis.  Examples of \hi absorption features are indicated  by the downward arrows in the top panel of Figure~\ref{fig:distance}. Note  that for illustrative purposes only a few selected dips are marked in the figure. Furthermore,  only those dips which lie within the velocity range of the \cii profile   are relevant to our analysis.      We identify the \hi absorptions without bias using an IDL algorithm ``EXTREMA'' \citep{meron1995}.
 Thus we were able to place  6510    at the near--distance  using the presence of \hi absorption. Identifying HISA features requires care as it is probable that some of the identified \hi absorption dips are not true HISA but represent some large scale feature.  Likewise we may miss some of them due to beam smearing. It is difficult to assess individually the reality of all the HISA features.  However because we are using a large ensemble in our analysis,  the inclusion of a small fraction of spurious features will not affect our results in a significant way.  Indeed, the fraction of HISA identified in our sample is consistent with the expectation for the frequency of HISA.  For example,  the fraction   in our sample ($\sim$61\%)  is roughly consistent with the frequency of HISA in the sample used by \cite{Duval2009}. In their survey of 702 $^{13}$CO  sources with near-- \space and far--distance ambiguity  they constrained the solutions to the near--distance  for $\sim$70\% of the sources.  For comparison, in our analysis of 3361 spaxels detected with $^{12}$CO we constrain $\sim$62\% of them to the  near--distance solution.

\subsubsection{2-D Galactic maps}
\label{sec:spaxel_2Dmaps}

We use the spaxel intensities and distances, in all latitude ($b$) data,  to produce 2-D Galactic maps of four gas tracers.   Figure~\ref{fig:2-D_map} shows the maps of Galactic \cii emission compared with \hi and CO emissions derived from the intensities in the spaxel and using the radial velocity distances. For \hi we use all spaxels, but for \cii and CO we use only the spaxels with 3-$\sigma$ detections.   The logarithmic spiral arms \cite [] []{vallee2008,Steiman2010} are plotted on the 2-D Galactic maps. Figure~\ref{fig:2-D_map} is the first 2-D representation of the Galactic \cii emission in the plane. We can see that most of the \cii emission is closely associated with the spiral arms and it is brightest near the spiral arms' tangent points.  In the Galactic intensity maps we can identify two major artifacts, both resulting from the $V_{lsr}$--Distance transformation.   The first are the radial striations (or ``fingers'') caused by velocity dispersions (peculiar or streaming motions) that confuse   the distance estimates resulting in  a   smeared radially elongated feature   along the line of sight  \cite[cf.][]{levine2006,englmaier2011}. For example, the elongated strips near the Galactic center are likely to be caused by such velocity dispersions. The second are the voids  seen near the tangent points in both longitude quadrants in all maps which are an artifact resulting from the algorithm used to resolve the near-- \space far--distance ambiguity. In the $V_{lsr}$--Distance solution  for spaxels with velocities close to the tangent velocity (Figure~\ref{fig:distance}b) the far--distances, up to within 1 kpc from the  tangent point,  are assigned to the near--distance. Thus the lack of spaxels at distances up to 1 kpc beyond the tangent point results in the void seen in the maps. Furthermore, all spaxels with $V_{lsr}$ exceeding the tangent velocity (due to peculiar or streaming motions)  are set to be at the tangent distance, resulting in an accumulation of spaxels at the tangent points. However, the scale of the radial striations or the void ($\sim$1 kpc) is within the distance uncertainty assumed for the analysis and therefore do not change our results for $z$-scales derived using the kinematic distances.   We do not discuss further the Galactic distribution of \cii in the spiral arms because it can be studied better using the HIFI OTF scan maps observed from other {\it Herschel} programs.  Another striking difference between \cii and CO  maps is the diminished \cii emission within 2 -- 3 kpc of the  Galactic center.   While  CO surveys find strong emission in the center \cite[cf.][]{dame2001}, in the BICE \cii data \cite{Nakagawa1995} reported a deficiency of \cii toward the Galactic Center.  While \hi is detected to large radial distances covering the full extent of the map the \cii and CO are not detected at radial distances  much beyond the solar circle, which is consistent with the radial distributions reported in \cite{Pineda2013}.


\section{Analysis of the ISM phases traced by \cii}

In the previous sections we described a method to represent the \cii survey statistically in terms of spaxels and discussed how we determine their location in the Galaxy.  We summarize the details of the  data used in our analysis   in  Table~\ref{tab:spaxel_analysis}. We identified a total of  23229 spaxels (each with a 3 \kmss wide velocity bin)   in the GOT C+ data selected by their \hi detections (3-$\sigma$ = 7 K \kms).  Of these only 6484 spaxels were detected in \cii (3-$\sigma$ = 0.5 K \kms) and 4421  in $^{12}$CO (3-$\sigma$ = 3.0 K \kms). Thus the vast majority ($\sim$72\%) of  \hi spaxels do not have detectible \cii emission, even though \cii emission is expected to be associated with both \hi and \h2 (the collisional excitation rate coefficient for exciting \c+ is about 1.5 times larger for H atoms than \h2 \cite[][]{wiesenfeld2014}). This statistical result on the association of \cii emission with \hi and CO raises the question, why is there such a low fraction of \hi associated with \cii emission, given the rather widespread distribution of \hi throughout the Galaxy?  Here we use the spaxel data set and the GOT C+ Gaussian fits \citep{langer2014_II}  to compare statistically their relationships and identify the  origin of \cii from different ISM components.

 As seen in the intensity distribution maps in Figure~\ref{fig:2-D_map},  the spaxels at Galactocentric radii R$_G$ $>$ R$_{\odot}$  are dominant only in \hi   with little or no \cii emission.  Therefore, for a  statistically significant comparison of the \cii detections with \hi it would be unrealistic to include the spaxel population at R$_G$ $>$ R$_{\odot}$, because it would bias the result.  Therefore we exclude these from further analysis.

\subsection{\ciis, CO Detections: Galactic distribution and Comparison with \hi}

\subsubsection{\hi Intensities/Column densities}
\begin{figure}[!ht]
\includegraphics[scale=0.52,angle=0]{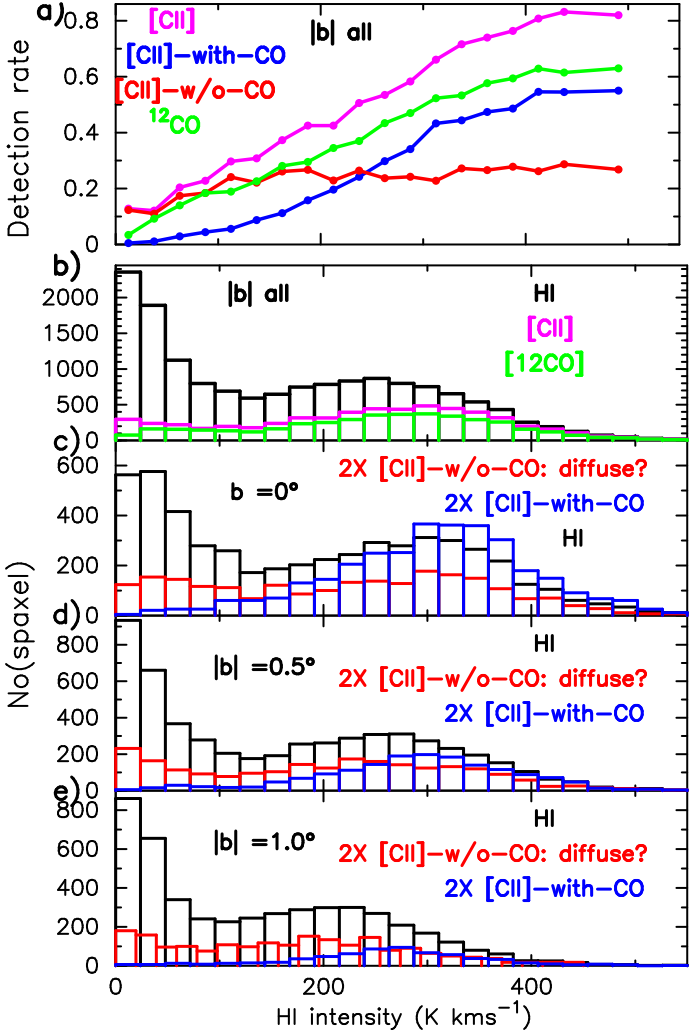}
\caption{Comparison of the gas components of \cii emission as a function of  \hi intensity for Galactocentric radius R$_G$ $<$ R$_{\odot}$ (8.5 kpc); and for different values of $b$ as indicated in panels (a) to (e) (see Table~\ref{tab:spaxel_analysis}). (a) The fraction of detections of \cii and $^{12}$CO components and combinations of \cii and CO (as indicated by the colors) plotted against \hi intensity. (b)  Histogram representation of the   number distribution of spaxels detected in \his, \ciis, and $^{12}$CO as a function of  \hi intensity.  (c) -- (e)   The number of \cii detections compared with the  \hi intensity   shown for each absolute value of the latitude separately.   The \cii detections with and without CO  are plotted separately   to bring out differences between  the dense and diffuse components of \cii emission.}
\label{fig:histogram}
\end{figure}

 In Figure~\ref{fig:histogram} we summarize the \cii and CO spaxel detections as a function of \hi intensity  for the 15278 spaxels identified at  R$_G$ $<$ R$_{\odot}$.   In panel Figure~\ref{fig:histogram}(a)  we display the fraction of spaxels with \cii and CO detections as a function  of  \hi intensity; this fraction is a measure of the detection rates of \cii and CO in the Galaxy. The spaxel fractions for each \hi intensity bin are  defined as the ratio of number of spaxels with detections of a given gas component (e.g., \ciis, \ciis-with-CO) to the total number of \hi spaxels in the bin. Both \cii and CO have higher detection rates at higher \hi intensities (corresponding to higher \hi column densities  and/or local H-atom densities).  The detection rates for \cii spaxels associated with CO show a nearly identical increase with \hi intensity as $^{12}$CO with \his, whereas the \cii spaxels without CO do not show any significant  increase with \hi intensity.  To further  examine  the fraction of detections  in Figure~\ref{fig:histogram}(a), we show  in Figure~\ref{fig:histogram}(b) the number distribution of \his, \ciis, and   $^{12}$CO  spaxels as a function of \hi spaxel intensity.   We find that for  R$_G$ $<$ R$_{\odot}$ a low fraction ($\sim$27\%) of \hi spaxels have CO detections and have an even lower fraction of  \cii with CO detections ($\sim$17\%).   We interpret this result as a consequence of the fact that only a small number of \hi clouds are seen in association with dense  \h2 gas.     Although the \cii  detection rates, especially for those with CO, correlate well with the \hi intensity, the \cii intensities do not show any correlation with \hi intensity \cite[see Fig. 9 in][]{langer2014_II} and show a large scatter (over two orders of magnitude). Therefore, we interpret the higher rate of detection at large \hi intensities, not as arising from the \hi gas itself, but  with the  probability for the presence of high density molecular gas associated with increasing \hi column density.  In other words, as \h2 fractions are expected to increase with increasing \hi column densities \cite[e.g.,][]{sternberg2014} the likelihood of detecting \h2 excited \cii emission also increases.

 About 17\% of the \hi spaxels are associated with both \cii and CO emission,  while a similar fraction  $\sim$20\%   of the \hi is associated with \cii emission without CO. To understand better the statistics of \hi versus \cii emissions    we show  in Figure~\ref{fig:histogram}(c -- e) the results for the GOT C+ LOS grouped by the absolute value of Galactic latitude. Note that here we plot the latitudinal  distributions for \cii detections with and without CO separately to bring out the remarkable differences in their distribution in and out of the plane.  We classify the \cii detections in the spaxels as (i) ``\ciis-with-CO'':  spaxels in which both \cii and $^{12}$CO emissions are detected, and (ii)  ``\ciis-w/o-CO'':  spaxels in which only \cii is detected and no $^{12}$CO emission is detected. The  overall number distributions of spaxel detections in Figure~\ref{fig:histogram}  show:

\begin{itemize}

\item Figure~\ref{fig:histogram}(a): The detection fractions (defined as the ratio  of the number  of  detections to the total number of \hi spaxels in a given \hi intensity bin) for \ciis-with-CO and $^{12}$CO show similar increase with the \hi intensity; however the detection fractions for \ciis-w/o-CO show   little or  no dependence on \hi intensity.
\item Figure~\ref{fig:histogram}(b): At intermediate and high \hi intensity both the number of \cii and CO detections (spaxels) show some degree of correlation with the total number of \hi spaxels.   However, at low \hi intensity, their detection rate does not show any relation to the number of \hi spaxels  at any given intensity.
 \item Figure~\ref{fig:histogram}(c--e): At  \hi intensities $>$200 K \kmss the number of \ciis-with-CO detections  shows  a stronger dependence on the number of \hi spaxels than that for \ciis-w/o-CO.  At lower \hi intensities ($<$200 K \kms) the \cii with or without CO detections are nearly independent of the number of \hi spaxels  in each intensity bin.
\item Figure~\ref{fig:histogram}(c--e): \ciis-with-CO detections  show a stronger dependence on Galactic latitude with the highest detection rates at $b$=0\deg, decreasing sharply as $|b|$ increases.
 \item Figure~\ref{fig:histogram}(c--e): In contrast to the  \ciis-with-CO,  the \ciis-w/o-CO detections  have (i) a very similar number distribution in all three latitudes and (ii) show no dependence on the number of \hi  spaxels in each \hi intensity bin.
     \item There is no apparent correlation between the \cii and \hi intensities in the spaxels with \cii detections (see below).
\end{itemize}

 In the following we use the results of the spaxel detection only to quantify the fraction of Galactic \cii likely to be associated with \h2 gas (see Section 4.2)  and to  differentiate it from the emission in the diffuse atomic \hi gas or the WIM (see Section 4.3).  Here we do not  calculate the amount of CO-dark H$_2$ in the FUV illuminated layers of CO clouds, which has already been analyzed  in our earlier work  \citep{langer2010,velusamy2010,velusamy2013,langer2014_II}.

\subsubsection{Galactic radial distribution}

\begin{figure}[!ht]
\includegraphics[scale=0.4,angle=-90]{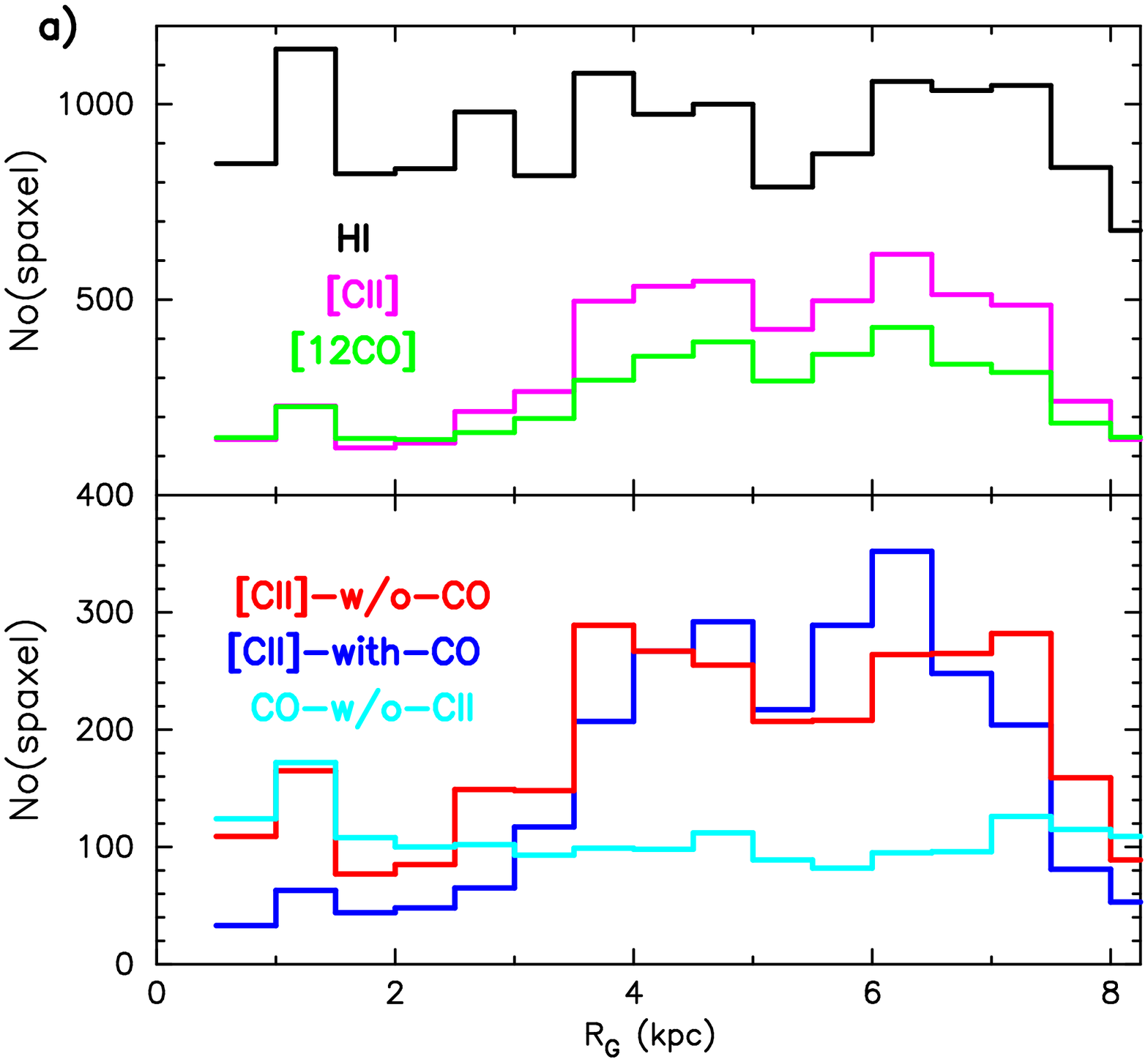}
\includegraphics[scale=0.4,angle=-90]{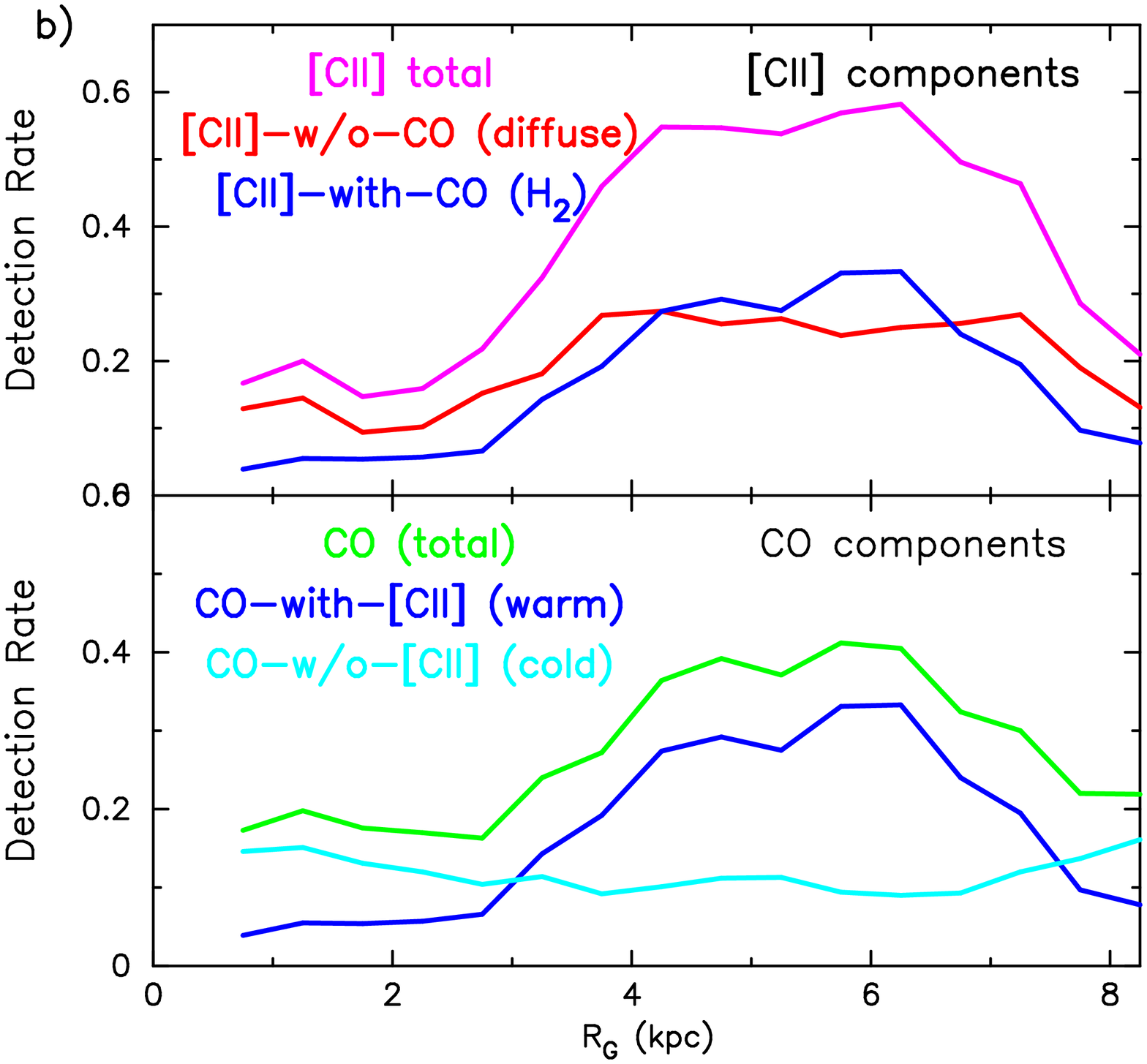}
\caption{   Comparison of the ISM gas components  as a function of  Galactocentric radius R$_G$. The spaxels for R$_G$ $<$ R$_{\odot}$ (8.5 kpc) are the only ones used (see Table~\ref{tab:spaxel_analysis}) for the analysis.   (a)  Histogram representation of the   number distribution of spaxels detected in \his, \ciis, and $^{12}$CO as a function R$_G$ (upper panel).  Number of detections of the \cii and CO components (lower panel) (b)  Detection rates of \cii (upper panel) and CO (lower panel) components (normalized using the number of  \hi spaxels in each R$_G$--bin as measure of the sampled volume) as function R$_G$. }
\label{fig:R_histogram}
\end{figure}

  The radial distribution of gas tracers gives the global characteristics of the ISM in the Galaxy  and are more robust without the ambiguity of the near-- \space far--distance determination discussed above.  The characteristics of the \cii emission, including the CO-dark H$_2$ gas, as a function of Galactocentric radius, R$_G$, in the Galactic plane have been reported in previous GOT C+ publications \citep{Pineda2013,langer2014_II}.  Here we present a slightly different perspective as revealed in our spaxel  spectral analysis of the GOT C+ data.   We use these Galactic radial distances of the spaxels and the spectral line detections in them to derive their distributions as a function of R$_G$. The spaxel data are binned in R$_G$ in 0.5 kpc widths sampled every 0.5 kpc.  We count the number of spaxel detections  and their total  intensities  in each bin for \his, \ciis, and $^{12}$CO. In Figure~\ref{fig:R_histogram}a (upper panel) are shown the  radial distributions of the number of detections of \his, \ciis, and CO  and in Figure~\ref{fig:R_histogram}a (lower panel) are shown the radial distributions of their individual gas components \ciis-with-CO, \ciis-w/o-CO, and CO-w/o-\ciis.

  The number of spaxels in each R$_G$--bin depends on the Galactic volume sampled. Because the GOT C+ survey does not sample the Galactic volume uniformly we need a ``volume'' normalization for each R$_G$--bin to interpret these distributions. We   use the total number of \hi spaxels in each R$_G$--bin for the normalization. As seen in the examples of the spectral line profiles of \cii and all ancillary data in the GOT C+ data base (see Figure~\ref{fig:data}) the \hi profile always has the broadest velocity range and encompasses all the emission features in \cii and CO. In other words, all the \cii and CO features lie within a, generally, broader \hi emission profile as a function of velocity. Thus all the spaxels in our sample have \hi detections. We can therefore use the number of \hi spaxels in a given R$_G$--bin to represent a measure of the  Galactic volume sampled within that R$_G$--bin. We define the detection rates of \ciis, CO and their gas components as the ratio of the number of detections in each R$_G$--bin of the respective gases to the total number of \hi spaxels in the R$_G$--bin.
In Figure~\ref{fig:R_histogram}b are shown the radial distributions of the detection rates, which are now free from the bin-to-bin  variations in the sampling, of the various \cii gas components (top panel) and CO gas components (lower panel).

\begin{figure}[!ht]
\includegraphics[scale=0.4,angle=-90]{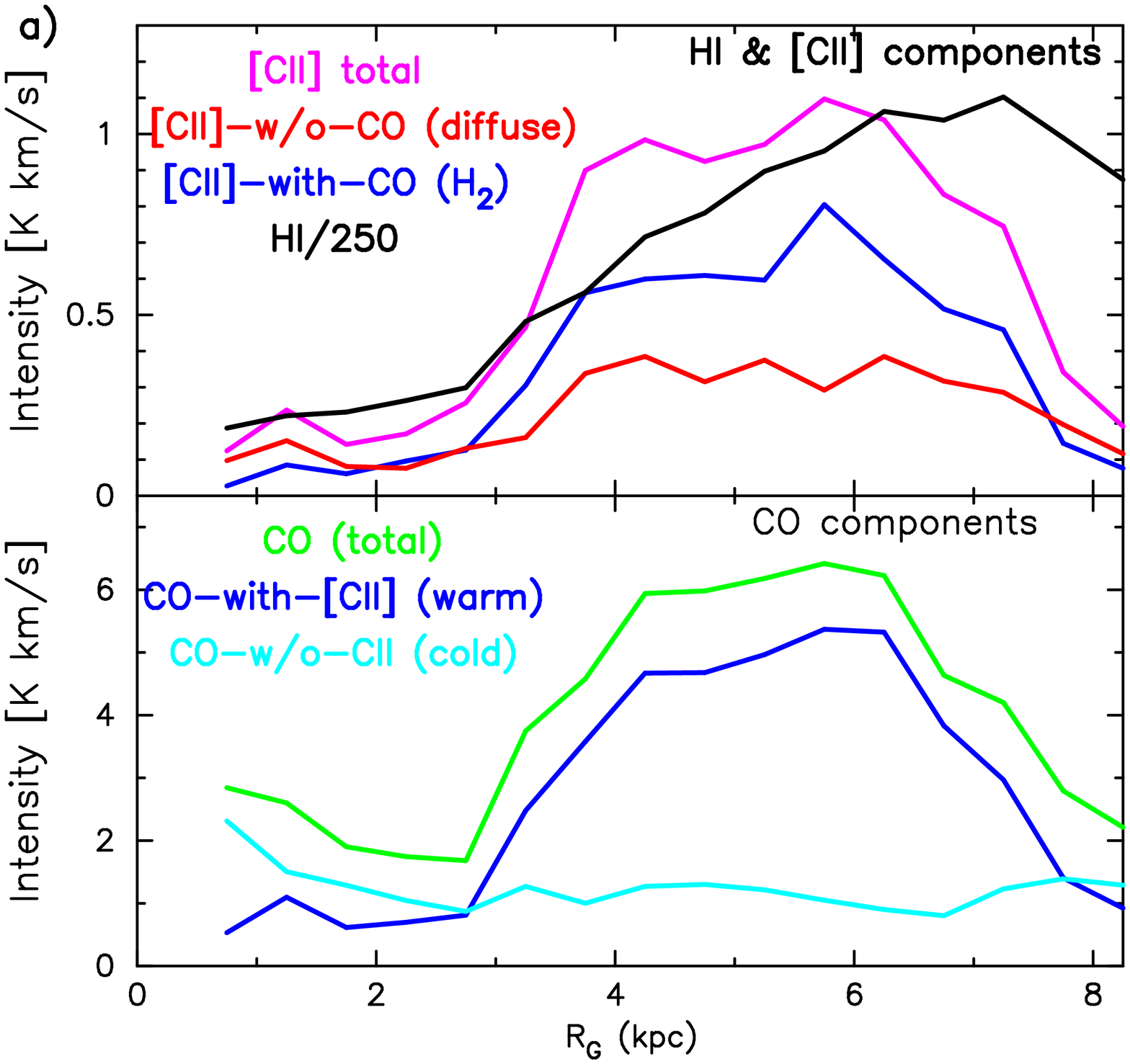}
\includegraphics[scale=0.4,angle=-90]{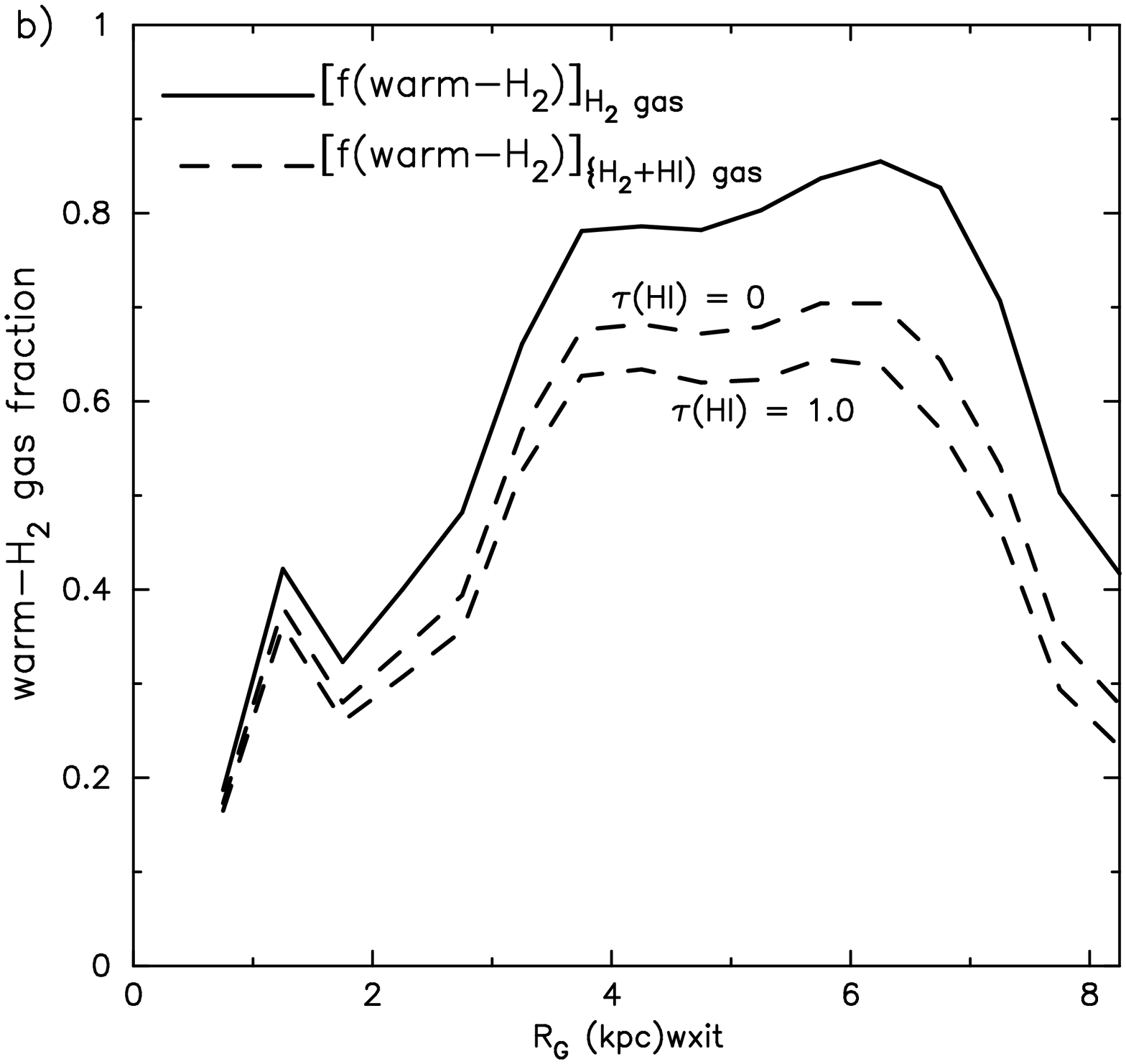}
\caption{ (a) Radial distribution of the spectral line intensities (normalized using the number of  \hi spaxels in each R$_G$--bin as measure of the sampled volume) of \his, \cii and its gas components (upper panel) and CO and its gas components  (lower panel). (b) Galactic distributions of warm \h2 gas fraction,  f(warm-\h2): (a)  with respect to the  total \h2 gas (solid line); (b) with respect to the total \his+\h2 gas (dashed line);    $\tau$(\his)=0 and 1.0 represent f(warm-\h2) for optically thin and thick cases used for \hi column densities, respectively.  The variation of the  \h2 fractions  as a functions of    Galactocentric radius R$_G$ are representative of the  star formation rate (SFR) and  star formation efficiency (SFE).  }
\label{fig:R_sfe}
\end{figure}

  We find that the \cii detection rates  peak in the range 4 kpc $<$ R$_G$ $<$ 7 kpc. The \cii with CO (the H$_2$ gas) is strongly peaked over this range of R$_G$, while \ciis-w/o-CO (the diffuse gas) has a shallower peak and a broader distribution in R$_G$. The radial distribution of CO is roughly similar to that of \cii with a broad peak in the range 4 kpc $<$ R$_G$ $<$ 7 kpc. {\it However, the CO gas components with and without associated  \cii   emission show distinctly different radial distributions}. The CO emission spaxels without \cii detections represent cold CO with H$_2$  gas clouds in which the C$^+$ layers are too cold for efficient excitation of the C$^+$ $^2$P$_{3/2}$ state. On the other hand the spaxels with detections of \cii along with CO represent H$_2$ clouds in which the C$^+$ layers are sufficiently warm ($>$ 35 K) for efficient \cii excitation.  Thus the results in the lower panel in  Fig.~\ref{fig:R_histogram}b delineate remarkably well the Galactic radial distributions of the warm--H$_2$ and cold--H$_2$ gas clouds. In contrast to the warm-H$_2$ broad peak in the range 4 kpc $<$ R$_G$ $< $ 7 kpc, the cold-H$_2$ shows a flat and more uniform distribution at all R$_G$.  Indeed there is even hint of a decrement in cold--H$_2$ near the peak of the warm--H$_2$ gas.

For quantitative analysis, as done above for the detection rates,   we derive
 ``volume'' normalized intensities of \ciis, CO and their gas components defined as the total spaxel intensities divided by the number  of  \hi spaxels     in each R$_G$--bin. In Fig.~\ref{fig:R_sfe}a are shown the radial distributions of the  intensities per unit volume  for \his,   \cii and its gas components (upper panel) and CO and its gas components (lower panel).    These results are broadly consistent with those derived by \cite{Pineda2013} using the azimuthally integrated intensities and the geometric area of each annular ring. The results presented here provide a good representation of the non-uniform sampling within the annuli but normalized to arbitrary volumes that may differ from spaxel--to--spaxel.  Thus, they  bring out clearly, and provide additional confirmation of, the global \cii and CO emission characteristics discussed by \cite{Pineda2013,Pineda2014}. Most importantly,
the differences in the radial  distributions for the cold and warm \h2--gas, as indicated by their association with or without \ciis, raise the possibility of using the intensities as  a measure  of the star formation efficiency (SFE) and/or rate (SFR).  Star formation will heat the gas and thus the warm clouds are signatures of recent star formation. The cold clouds, in contrast, are located in regions without recent star formation. Therefore by combining both \cii and CO emission in the spaxel analysis along with \his, we have a measure of the SFR and/or  SFE.

We assume that the CO intensities in the spaxels measure the H$_2$ column density (neglecting the CO-dark H$_2$ in the C$^+$ layer). Although the $^{12}$CO emission is generally optically thick it has been shown that the intensity of $^{12}$CO is proportional to the column density of H$_2$ of molecular clouds \cite[see the recent review by][]{Bolatto2013} and N(H$_2$)=X$_{\rm CO}$I($^{12}$CO), where X$_{\rm CO}$ is the CO-to-H$_2$ conversion factor. While the cold--H$_2$ is found to be roughly uniform with R$_G$, the warm--H$_2$ is predominantly in the range 4 kpc $<$ R$_G$ $< $ 7 kpc. The presence of warmer H$_2$ is an indication of a star formation environment. Thus,   in Fig.~\ref{fig:R_sfe}(b) we plot the warm--\h2 gas fractions with respect to (i) the total \h2 gas (solid line) and (ii)  total gas including \hi (dashed lines).    We use the \hi and CO intensities in each R$_G$--bin to derive their column densities, N(\his) and N(H$_2$).   For \hi column densities we consider   optically thin ($\tau <<$ 1) and thick ($\tau$ = 1.0) cases for the inner Galaxy \cite[e.g.,][]{Kolpak2002}  as marked in Fig.~\ref{fig:R_sfe}b.  The \hi column densities are corrected for the optical depth using Eq. 8 in \cite{chengalur2013}.   Treating the presence of warm--H$_2$ as evidence for star formation we can relate the radial profile warm--\h2 gas fractions   to  that of   the star formation rate and/or efficiency.  The radial profiles  of the warm--\h2 gas fractions with respect to total gas    have a flat peak   (value between 0.6 and  0.7) at R$_G$   between 4 and 7 kpc.    The star formation rate will depend on the star formation efficiency and the gas available to convert to stars, and the spaxel analysis used here provides an important constraint on the warm-\h2 gas fractions and thereby the radial profile of star formation in the Galaxy.

\subsection{\ciis-with-CO emission tracing the dense \h2 gas \\}

The  apparent  correlation of the \cii detection rate with \hi intensity in Figure~\ref{fig:histogram}(a) agrees so well with that for CO with \his, that we can, with confidence, partly attribute some of the \cii emission to a dense \h2 gas component, and it is an indication that the \cii does not originate primarily in the \hi gas. It is further evident by the fact that at high  \hi intensities the detections of \ciis-with-CO show a similar dependence  on \hi intensity as CO,   while that for  \ciis-w/o-CO show little dependence   on the \hi intensity.   This correlation  shows that both \cii and CO have higher detection rates at higher \hi intensity, but their intensities do not correlate with \hi intensity. This result can be interpreted as follows: the higher \hi intensities  mean higher \hi column density within the spaxel volume, thereby, increasing the probability  for the presence of \h2 gas in it. This is consistent with the fact that the \h2 gas fraction increases with \hi column density \cite[e.g.,][]{braine2011,braine2012,sternberg2014}. 
Thus  we   conclude that for all spaxels having \ciis-with-CO   the significantly large \cii is excited by \h2 gas, with only a small contribution  coming from the   \hi layers  \cite[cf.][]{langer2014_II}).  We    estimate the contribution from the \hi layers as follows.

The  \hi column density N(\his) [cm$^{-2}$] in each spaxel, in the optically thin limit, is proportional to the \hi intensity: N(\his) = $1.82 \times 10^{18}$ I(\his) cm$^{-2}$ where I(\his) is in units of [K \kms].  We  derive the column density of \c+  from N(\his) using the procedure given  in  \cite{langer2014_II} [their Equation A5] and the assumptions therein, N(\c+) =  N(\his) $X_H$(C$^+$) where  $X_H$(C$^+$)  is the carbon abundance.  Here we assume  $X_H$(C$^+$) $\sim$2.2$\times$10$^{-4}$,  typical for the Galactocentric radii where most \cii arises. Expressing the \hi density and temperature in terms of a pressure (P = n(\his)$\times$T$_K$  K cm$^{-3}$) and \hi intensity in the spaxel (I(HI) (K \kms) we  can write the emission in the \hi gas simply as,

\begin{equation}
I(CII)_{HI} = 1.03\times[P/3000]\times[I(HI)/1000]  {\rm {\, [K\, km\, s^{-1}]}}
\end{equation}

From the analysis above we conclude  that all \ciis-with-CO detections, which trace the dense molecular clouds, are associated with  \h2 gas, and that   their  \hi layers make only a small contribution to the spaxel \cii intensity.  We  compute the emission from the \h2 gas as I(\ciis)$_{H_2}$ = I(\ciis)$_{total}$-- I(CII)$_{HI}$.    We estimate  the contribution from \hi gas, assuming  a \hi gas pressure, P $\sim$    5000 K cm$^{-3}$ (corresponding to the median spaxel radial distance,  R$_G$ $\sim$ 5.5 kpc \citep{wolfire2003,Pineda2013}).  The fractional emission from these dense \h2 clouds is   given later in the final summary   in   Figure~\ref{fig:summary}.

\subsection{\ciis-w/o-CO  emission: as diffuse \h2,  diffuse  \his, or WIM?\\}

 It is evident in Figure~\ref{fig:histogram}(a,c--e) that  for the \cii detections with CO  there is some correlation with \hi intensity, whereas for those without CO  there is no direct correlation   with the number of \hi spaxels or the \hi spaxel intensity. Therefore,  for the analysis  of the diffuse \cii emission   we   exclude all the spaxels with CO detections as they are definitely associated with the dense \h2 gas.  Thus the \cii detections in the spaxels without associated CO detections (designated as
 \ciis-w/o-CO) represent emissions originating  in (i) diffuse  \h2 gas with no CO (or diffuse CO--faint \h2 gas), (ii) \hi clouds  and  (iii) the   WIM.

\subsubsection{Gauss fit data: (CO-faint) diffuse \h2 gas\\}

\begin{figure}[!ht]
\includegraphics[scale=0.35,angle=-90]{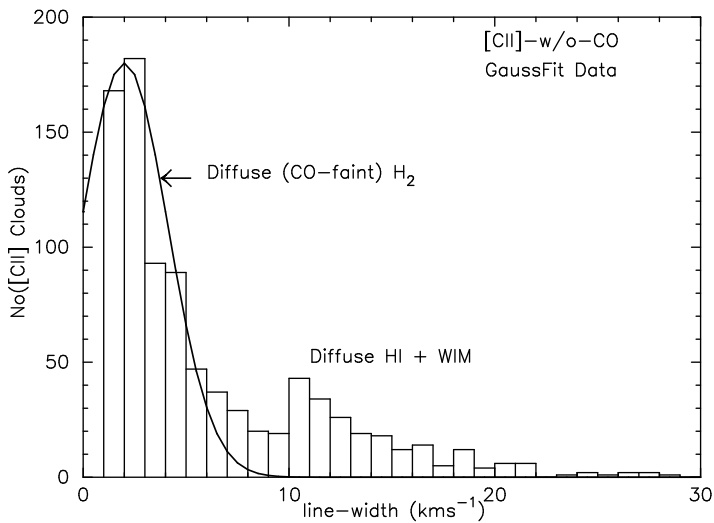}
\caption{A histogram of the Gaussian line widths of  \cii velocity components without CO in the GOT C+ data.  The distribution is used to delineate the broader components, corresponding to low surface brightness extended emission arising from diffuse \hi or the WIM, from the narrower line width associated with the diffuse molecular (CO faint) \h2 gas.   The Gaussian line widths are taken from the data base assembled by \cite{langer2014_II}.}
\label{fig:diffuse_H2}
\end{figure}
 An additional component of \h2 molecular gas  which is missed by CO but detected in \cii has been  identified in the GOT C+ survey \cite[cf.][]{langer2010,velusamy2010,langer2014_II}.  This component represents an early phase of cloud evolution or a Galactic environment where little or no CO exists because the gas is suffused with photodissociating UV radiation. Unlike  the situation found for  the spaxel data it is easier to identify such  diffuse molecular clouds in the  GOT C+ Gauss fit spectral line profile data base using the line width to distinguish  between the diffuse molecular (as narrow line widths) and diffuse \hi or WIM (as the  broader line widths, corresponding to line wings in Figure~\ref{fig:linewings}).     To analyze the diffuse gas we  exclude all Gauss fit \cii components with associated CO emission. Then we examine the velocity widths of the remaining Gaussian fitted components to separate the denser \h2 clouds from the diffuse gas.  Figure~\ref{fig:diffuse_H2} is a histogram plot of the number of \cii spaxels as a function of line width for these \ciis-w/o-CO components.  There is a narrow distribution of linewidths ($\Delta$V  $<$ 7 or 8 \kms) that peaks at $\sim$3 \kms.  This line width  range is characteristic of dense molecular clouds.  However, there is a tail to the distribution with a peak $\sim$10 to 15 \kms. These  broader line widths are similar to the  spectral line ``wings'' shown in Figure~\ref{fig:linewings}  and can be interpreted as diffuse \hi gas or WIM.  Note this \h2 gas component is different from the CO-dark \h2 gas in the outer layers of CO clouds. Following \cite{Bolatto2013} here we refer to it as (CO-fant) diffuse \h2 gas. (There is some confusion in the literature as there is no agreed upon label for this component, and it  is also referred to as CO-dark  \h2 gas \citep{Pineda2013}, diffuse CO-dark H$_2$ gas \citep{langer2014_II},  ``CO faint" molecular \h2 gas \citep{Bolatto2013}, and dark gas \citep{grenier2005,wolfire2010}.)  Using the Gaussian fits for the \ciis-w/o-CO cloud components in the GOT C+ data base of \cite{langer2014_II},   we estimate the sum of the \cii intensities for    (i) the narrow   ($<$ 8 \kms) velocity width components (seen in Figure~\ref{fig:diffuse_H2}) as a measure total intensity of the diffuse \h2 gas,   (ii)  the broader   ($>$ 8 \kms) velocity width components   as a measure total intensity of diffuse  atomic \hi clouds and/or WIM gas, and  (iii) all the Gauss fit components with or without CO as a measure of the total Galactic \cii emission.   Then we   estimate the fraction  of the Galactic \cii emission in the  diffuse \h2 gas and that in the atomic \hi  and/or WIM.    (The  results are  also shown in the final summary  in   Figure~\ref{fig:summary}.)

\subsubsection{Spaxel data: How much of \ciis-w/o-CO is in \hi  gas? \\}

In the \ciis-w/o-CO spaxels data it is  difficult to separate the (CO-faint) diffuse \h2 gas from the diffuse \hi  or WIM, unlike in the Gauss fit data discussed above, where we identify individual features from their line widths. Here we  take a different approach which starts by examining the amount of \cii that we expect to arise from the \hi gas and use this criterion to distinguish whether the source is truly diffuse \h2.  The question, how much \cii emission is expected in the \hi clouds or the \hi gas layers surrounding the molecular clouds, has been been above in Section 4.2.  In brief, one uses the \hi intensities to derive the column density N(H) and scales it by the fractional abundance of ionized carbon to derive the column density of \c+, N(\c+) as in Eq. 1.  However, such estimates are highly model dependent as the C$^+$ excitation  has a high critical density, $\sim$3100 cm$^{-3}$ for atomic H, and a relatively large excitation energy, $\Delta$E/k = 91.2K.  Therefore the emission intensity is very sensitive to both density and temperature, or equivalently gas pressure \cite[see][]{langer2014_II}. Furthermore the uncertainty in metallicity used to derive the \c+ column densities from scaling the  \hi column densities is also a factor.

\begin{figure}[!ht]
\hspace{-0.5cm}
\includegraphics[scale=0.475,angle=0]{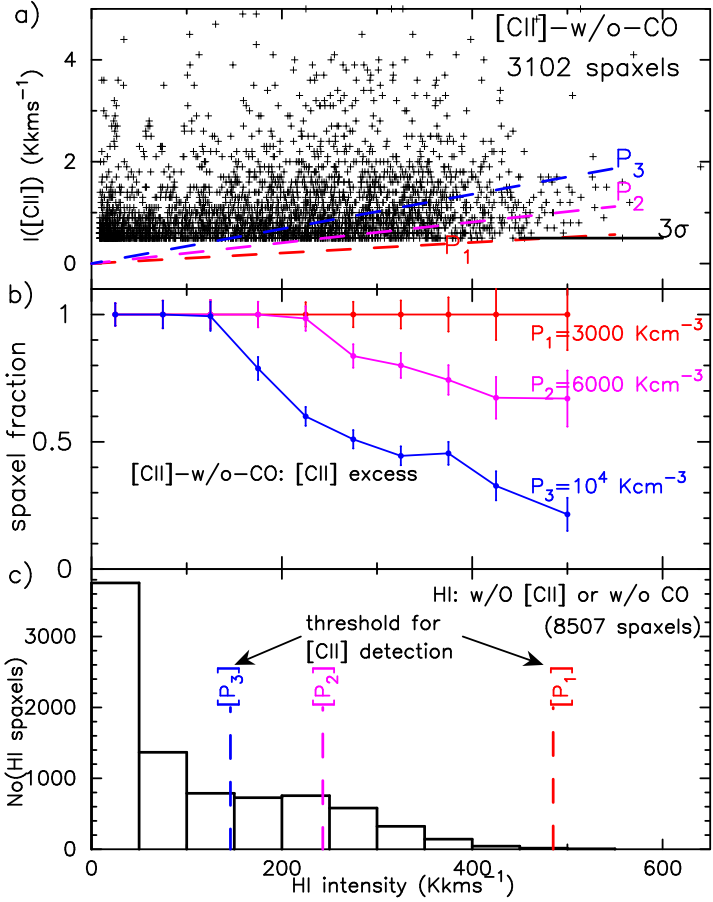}
\caption{Samples of diffuse \ciis-w/o-CO  and \hi without \cii or CO for R$_G$ $<$ R$_{\odot}$. The gas pressures  P$_1$,   P$_2$,  and P$_3$ in all panels (a) -- (c) correspond  to 3000,   6000, and 10$^4$ K cm$^{-3}$, respectively. (a) Scatter plot of \cii intensities verses \hi intensities for each spaxel.   The dashed lines  represent the \cii intensities expected for H excitation of \c+,  corresponding to the \hi intensity in each spaxel.  The solid lines represent the 3-$\sigma$ detection limit for \cii in the 3 \kmss wide spaxels. (b) Plot of the fraction of diffuse  \cii spaxels (\cii without CO) in each \hi intensity bin with \cii intensities in excess of that expected for excitation in the \hi clouds (see text).    (c)  Number distribution of \hi spaxels without \cii or CO plotted against \hi intensity for the sample of diffuse \hi spaxels without \cii or CO.    The dashed vertical lines indicate the minimum threshold \hi intensity for detecting \cii for pressure models P$_1$, P$_2$, and P$_3$.    }
\label{fig:cii_spaxels_noCO}
\end{figure}

We use Equation 1 (in Section 4.2) to estimate the \cii emission, I(CII)$_{HI}$, that can arise in the \hi clouds and or layers.
 In Figure~\ref{fig:cii_spaxels_noCO}(a) we plot the \cii spaxel intensity against the \hi spaxel intensity, I(\his), for all  \ciis-w/o-CO detections within R$_G$ $<$ R$_{\odot}$.  This  scatter plot contains no indication of any correlation between the \cii and \hi intensities. In this plot we also show (by dashed lines) a model calculation of the I(\ciis)$_{HI}$ intensity as a function of \hi intensity that is expected from the excitation of C$^+$ by the \hi gas within each spaxel volume for three assumed gas pressures: P$_1$ =3000 K cm$^{-3}$,  P$_2$ = 6000 K cm$^{-3}$, and P$_3$ = 10$^4$ K cm$^{-3}$.  \cite{langer2014_II} discuss in detail the appropriate gas pressures to estimate the \cii intensity in \hi gas. The lower pressure P$_1$  is more realistic for a typical \hi gas cloud in the solar neighborhood and P$_2$ is a maximum within the molecular ring \citep{wolfire2003,Pineda2013}, while the highest pressure P$_3$, which is typical of dense molecular gas, seem very unlikely for  the \hi gas layers or clouds \citep[][]{langer2014_II}.

If this model of emission in \hi clouds is correct then, as can be seen in Figure~\ref{fig:cii_spaxels_noCO}(a), for the low pressure case, P$_1$, none of the spaxels have detectable \cii from excitation by atomic hydrogen at the detection limit of the GOT C+ survey and all show \cii excess above what is expected for excitation by \hi (Figure~\ref{fig:cii_spaxels_noCO}b). If all the \cii intensity comes from the \hi gas we will not see any \cii excess and the fraction of spaxels with excess \cii will be zero. For the higher pressure, P$_2$, some spaxels should be detectable in \ciis, but even for P$_3$ the majority of the \hi spaxels are unobservable in \cii at the sensitivity of the GOT C+ survey. We illustrate this result in another way by plotting the spaxel fractions with \cii excess in each \hi intensity bin in Figure~\ref{fig:cii_spaxels_noCO}(b).  The \cii excess are computed as the observed \cii in each spaxel {\it minus} the  \cii intensity predicted for the \hi intensity and assumed gas pressure.  For the model with P$_1$ all have \cii excess and therefore the spaxel fraction is one for all \hi intensities.  For pressures P$_2$ and P$_3$ the spaxel fraction with \cii excess decreases for \hi intensities beyond $\sim$ 250 and $\sim$ 150 K \kms, respectively.   Even in the higher pressure cases   only a small percentage of the spaxels    have \cii intensities  that could be produced by \c+ excited by atomic hydrogen.  

What is the likelihood that we are underestimating the contribution to \cii emission from \hi if the observed \hi intensities are beam diluted or if there are optical depth effects?
 If the \hi is not optically thin then N(H) is underestimated and correspondingly the true \cii intensity is larger than  those  for the P$_1$ or P$_2$ curves in Figure~\ref{fig:cii_spaxels_noCO}(a).
  For example, \cite{Kolpak2002} estimate $\tau$ $\sim$ 0.8 to 1.4 for the inner Galactic regions between 4 and 8 kpc radius.  Thus for a gas pressure P$_1$ (which is more realistic) combined  with a factor of two intensity correction due to beam dilution  or opacity, the emission from the \hi gas   doubles and will  be identical to the predicted  intensity shown for P$_2$ without beam dilution  or optical depth corrections.  However, we cannot expect  beam dilution to be a significant effect much beyond this value, especially near the  highest \hi intensities,  because it will produce extremely high values of \hi intensities, resulting in very high column densities, N(H). However such high \hi column densities  are most likely to form molecular gas clouds, and will be detected as CO clouds with corresponding \cii emission.   Therefore  the spaxel fractions with \cii excess  can be represented realistically by pressures between P$_1$ and P$_2$.  Also, as shown below,  P$_3$ seem unrealistic for the \hi gas layers and we show it in Figure~\ref{fig:cii_spaxels_noCO}     to illustrate an extreme \hi pressure.

We can  also use our sample of   8507   diffuse \hi  clouds identified in the spaxels without \cii or CO detections, to check for the consistency of the above estimates  constraining  the amount of  \cii emission from H excitation of \c+ (that is, the number of spaxels below our \cii detection limit).  This sample  of spaxels clearly excludes any contamination from high density \h2 gas because none contain $^{12}$CO.   In Figure~\ref{fig:cii_spaxels_noCO}(c) are shown the intensity distribution  of these   \hi spaxels along with the threshold limits for \cii detection. In the case of P$_1$ all the spaxels  have \hi intensities below that required for detecting  \cii emission at the sensitivity of the GOT C+ survey.  However, for gas pressures P$_2$ and P$_3$, GOT C+ should have detected many of these  spaxels in \cii emission. Thus in these clouds only the low pressure case, P$_1$, is consistent with the observations of \hi   but no \cii detections. As seen in Figure~\ref{fig:cii_spaxels_noCO}(b), only a small fraction \hi spaxels have intensities consistent with excitation by H while the majority have a large \cii excess which cannot be accounted for by emission from \hi gas alone. This  excess could arise  from (CO-faint) diffuse \h2 gas through excitation by \h2 or from the WIM excited by electrons.

 \subsubsection{\cii fractions in (CO-faint) diffuse \h2 gas {\it versus} diffuse \hi or the WIM\\}

\begin{figure}[!ht]
\includegraphics[scale=0.45,angle=0]{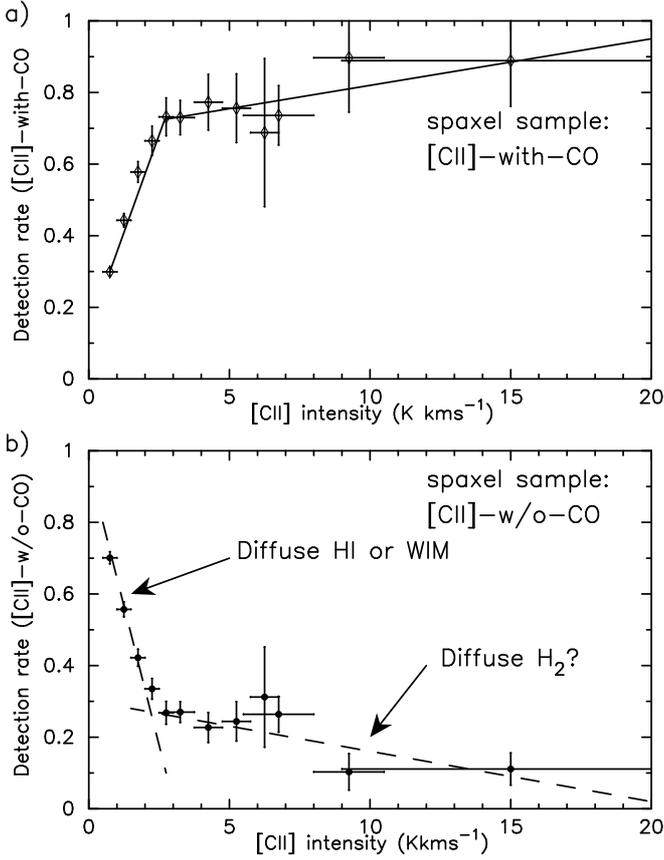}
\caption{The detection rates of  \cii gas components as a function of \cii intensity: (a) Spaxel sample for \ciis-with-CO  detection.  (b) Spaxel sample for \ciis-w/o-CO detection.    The error bars represent the widths of the \cii intensity bins and the  statistical error for each bin. The solid  lines are approximate fits to the detection rates as a function \cii intensity.   The labels in the lower panel indicate the corresponding \cii gas components.  }
\label{fig:CII_noCO_int}
\end{figure}

As seen in Figure~\ref{fig:cii_spaxels_noCO}(a) \& (b)   large number of spaxels in the \ciis-w/o-CO  sample   have a  \cii intensity excess  above that predicted from excitation in \hi layers and clouds,  and this fact strongly favors the  C$^+$ emission arising from the \h2 diffuse gas or WIM.  \cite{velusamy2012} suggested that excitation by electrons in the WIM  accounts for a similar   excess observed for the low surface brightness \cii along the Scutum-Crux spiral tangency.  Therefore, we  use the spaxel \cii intensity distributions to separate the emissions in the diffuse \h2 gas from that in the  diffuse \hi gas or the WIM.  We examine the fraction of \cii detections with and without CO as a function of \cii intensity.  The data are divided into \cii intensity bins and in each bin the number of \cii detections with and without CO are counted.  In Figure~\ref{fig:CII_noCO_int}(a) and (b) we plot the fraction of spaxels of \ciis-with-CO and \ciis-w/o-CO, respectively, as a function of the \cii intensity. (The fraction \ciis-w/o-CO is just one minus that with CO shown in Figure~\ref{fig:CII_noCO_int}a. However for convenience we plot it separately in Figure~\ref{fig:CII_noCO_int}(b).)  Here, 
 the value of spaxel fraction for the \ciis-with-CO   in a given \cii intensity bin is defined as   the ratio of the  number of \cii spaxels with CO detections   to the total number of \cii spaxels in the bin. Thus a fraction of unity will mean every \cii detection has a CO counterpart and zero means none of the \cii detections have CO.

 As seen in  Figure~\ref{fig:CII_noCO_int} the fraction of \cii spaxels with and without CO show distinctly different relationships with the intensity of \ciis, as denoted by the full and dashed lines.  The 
\ciis-with-CO spaxels are more dominant at higher \cii intensities, while the  \ciis-w/o-CO  
detections are dominant at lower \cii intensities. In the case of \ciis-with-CO it is not surprising because  we already know (see Table~\ref{tab:spaxel_analysis}) that this component is a dominant contributor to the Galactic \cii emission.  Of particular interest is the  break in the slope of the dashed line near $\sim$ 2.5 K \kms.  This break  identifies the following two populations:  (i) a set of \cii spaxels  with a detection rate of \ciis-with-CO increasing rapidly toward low \cii intensities, which we identify as diffuse \hi or WIM gas components and  (ii) a set of \cii spaxels  with a detection rate of \ciis-w/o-CO decreasing slowly with increasing \cii intensity, which we identify as a (CO-faint) diffuse \h2 gas component.   Therefore we can interpret the spaxels in the long tail in Figure~\ref{fig:CII_noCO_int}(b) at intensities $>$ 2.5 K \kmss as molecular gas and the lower intensity spaxels as the diffuse \hi or WIM gas.  Thus the break point in the \cii intensity provides  a useful marker for the sources of \cii in these gas components and we estimate the \cii gas fractions using the intensities above and below it.  The fraction of \cii in the diffuse \h2 gas without CO calculated from the spaxels  is in good agreement with that estimated from the Gauss-fit analysis  (Section 4.3.1).

\begin{figure*}[!ht]
\vspace{-0.25cm}
\centering
\includegraphics[scale=0.85,angle=0]{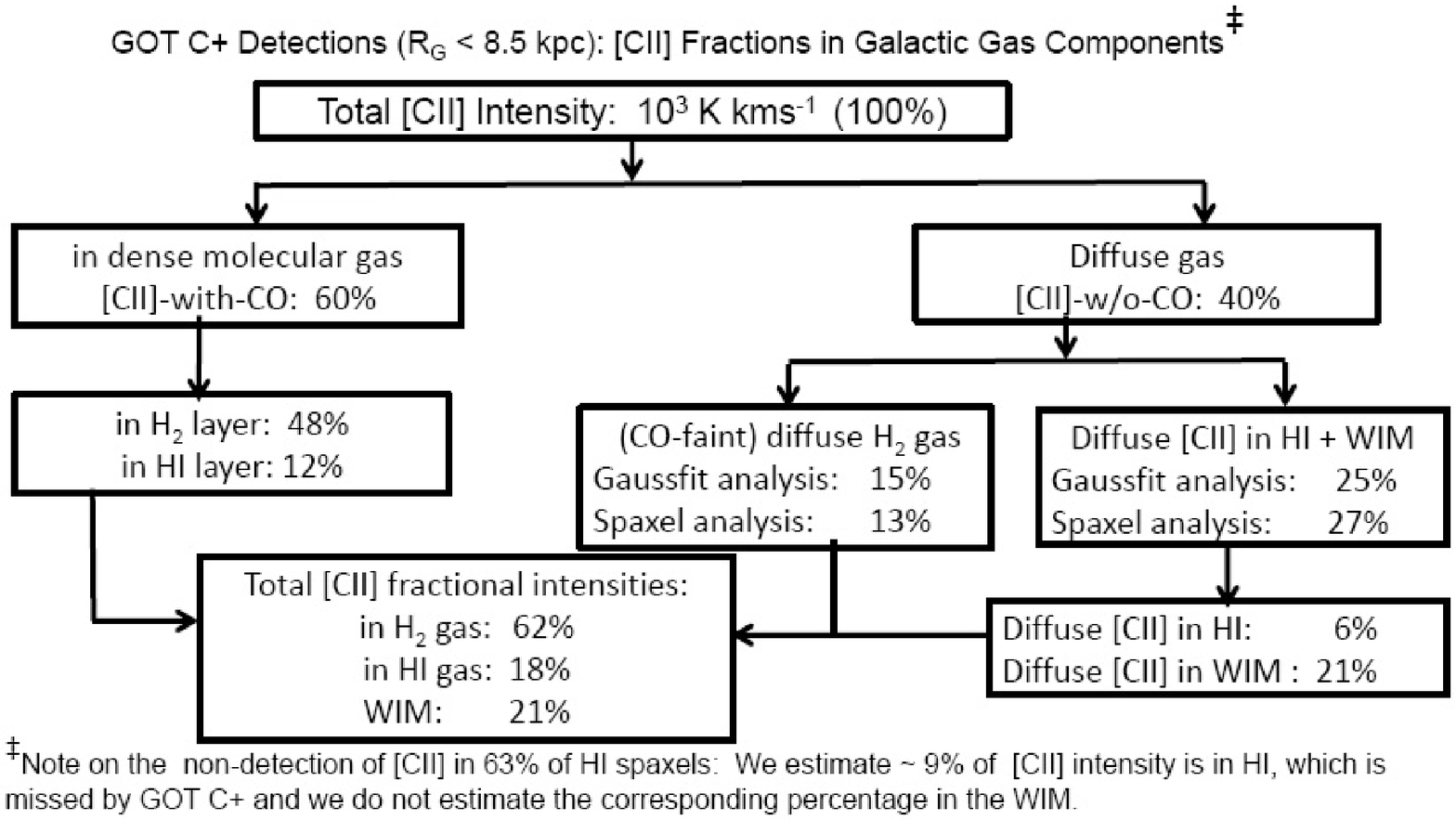}
\caption{Summary of the Galactic \cii emission fractions in different gas components in the inner Galaxy R$_G$ $<$ R$_{\odot}$.  Note here we include only the estimated fraction in \hi gas for  the non-detections of \cii in the \hi spaxels  but not the fraction  from the WIM (see text in Section 4.4). We also note  that the total   \cii gas fractions given here are estimated using the spaxel intensities alone and the estimates that include their z-scales are listed in Table~\ref{tab:galactic_fractions}}.
\label{fig:summary}
\end{figure*}

 In Figure~\ref{fig:summary} we show a flow  diagram of the separation of \cii emission into the different ISM gas components. As   shown  in Figure~\ref{fig:summary} from the GOT C+ spaxel analysis of \cii detections,  we obtain a \cii intensity fraction of $\sim$ 27\% from the diffuse \hi and WIM components combined. Assuming a pressure of 3000 K cm$^{-3}$ in the diffuse \hi clouds (see panel (c) in Figure~\ref{fig:cii_spaxels_noCO}) and their observed \hi intensities  we estimate  $\sim$ 6\% of \cii emission   comes  from the diffuse \hi gas. We can then assign the   remaining  21\% of the \cii intensity fraction to the WIM component. Thus in the diffuse ISM the WIM is more dominant as  a source of \cii emission than \hi gas with a ratio of \ciis$_{WIM}$ to \ciis$_{\rm {H I}}$ $\sim$3.   However,  \cite{Pineda2013} calculated only 4\% for the WIM   emissivity at $b$=0\deg, using a model for the electron density distribution. It should be noted that this model only reproduces the electron density distribution over large scales and so underestimates the density due to neglecting the small filling factor ($\sim$5\% to 10\%) for the WIM at $b$=0\deg.   As the intensity is proportional to the density squared \citep{velusamy2012} times the filling factor, $f$, the emissivity $\propto f \times n(e)^2$ and \cite{Pineda2013} underestimates the actual WIM emissivity. Finally, we note that at the sensitivity of the GOT C+ survey we can only detect \cii from the WIM in enhanced density regions,  $n(e)$ $>$  a few times the average electron density in the ISM such as those associated with the spiral arms \citep{velusamy2012}, thus weighting our estimate of the fraction of \cii from the WIM to more intense regions.  Our result that a larger fraction of \cii is in diffuse ISM is further corroborated by the broad $z$- scale distribution  as discussed in Section 5.2. However, our estimate for the WIM fraction in \cii is subject   to the uncertainty in the gas pressure and the metalicity  used for calculating the \cii intensity in the diffuse \hi gas (see Eq. 1).

 \subsection{Limit on \cii in \hi gas in spaxels with no \cii detections}

  The analysis so far,    and    the \cii fractions listed in  Figure~\ref{fig:summary},  only used   the gas components with detected \cii emission in their respective spaxels (above the 3-$\sigma$ limit).  For example, the 8507 \hi spaxels in the inner Galaxy (see Figure~\ref{fig:CII_noCO_int}(c)) are not included in the gas fractions as they all have \cii intensities below the detection limit.  Nevertheless their contributions to the total \cii intensity and to gas fractions although small in each spaxel, may be significant  in total.  We can use the models discussed in Section 4.3.2 for \c+ excitation by   atomic H  to estimate how much \cii intensity will be expected from this sample of \hi spaxels. From  Figure~\ref{fig:CII_noCO_int}(c) it is clear that only the gas pressure P$_1$ or less is appropriate for this sample.  Using Equation 1 we estimate for these \hi spaxels  a total \cii intensity  of 928 K \kms, which amounts to  $\sim$ 9\% of the total \cii intensity detected by GOT C+ in the inner Galaxy (see Table~\ref{tab:spaxel_analysis}).  However, we   cannot estimate the corresponding \cii emission in the WIM for the spaxels with \cii intensities below the detection limit without detailed modeling of the WIM in each spaxel. We do not have any observational data, such as Emission Measures (EM) and/or electron densities, relevant for quantifying the electron excitation in each spaxel,   unlike  in the case of diffuse \hi for which we could use the \hi intensity for quantitative analysis of excitation in the \hi gas.   Therefore, for comparison between the \cii fractions in \hi and WIM we should regard our estimate for the WIM as a lower limit, which still makes the WIM  a more dominant contributor to total  \cii emission than \hi.

If we include the above limit on \cii in the \hi gas in the spaxels with no \cii detection,  the \cii fraction in the diffuse \hi gas is increased to $\sim$ 14.5\% for cloud pressure P  = 3000 K cm$^{-3}$.  We estimate a similar amount, $\sim$12\%  of \cii in \hi gas associated with the denser molecular \h2 clouds (assuming a \hi gas pressure P  = 5000 K cm$^{-3}$ and  metallicity $\sim$ 2.2 $\times$ 10$^{-4}$ appropriate to the median spaxel radius R$_G$ $\sim$ 5.5 kpc).    Thus our estimate of a total  $\sim$27\% for \cii in \hi gas is close to those derived by \cite{Pineda2013} for the Milky Way and \cite{Mookerjea2011} for M33.    Therefore, we conclude that the total  observed \cii intensity originating from the \hi gas (see Figure~\ref{fig:summary})   is not the dominant source of the total of \cii emission when compared to that from the \h2 gas  or the WIM in the Galactic disk.


\section{$z$--distribution of \cii emission gas components\\}

The scale height for ISM components in the disk is important for understanding the pressure and energetics of the gas,   and for calculating the luminosity of the Galaxy in various gas tracers. In particular the determination of the \cii luminosity is important because it is routinely used to trace galactic star formation.   Whereas, those components traced by \hi and CO are well established from large scale Galactic surveys, the gas traced by \cii is not well established because the needed spectral line surveys of the 158-\microns line have not, until recently, been available.  Prior to GOT C+ all interpretations of the $z$-scale for \ciis, were based solely on $b$--scans without distance information.  These other large-scale Galactic plane surveys of \ciis,  namely,  COBE \citep[cf.][]{Wright1991,Bennett1994}, BIRT \citep{Shibai1991}, FILM onboard IRTS \citep[][]{Shibai1994,Makiuti2002}, and BICE \citep{Nakagawa1998} are spectrally unresolved.  Consequently the $z$--distribution of the clouds traced by \cii still remains  not well characterized.

  Only BICE, which surveyed the inner Galaxy (-350\degs $< l <$ 25\deg), had sufficient latitudinal coverage with reasonable angular resolution, 15\arcmin, to determine the distribution in $b$ up to the limit of their survey at $|b|$ = 3\deg.  FILM observed the \cii distribution to larger values of $b$  than BICE \citep{Makiuti2002}, but FILM only observed \cii in a narrow strip along a great circle crossing the plane at $l$ = 50\degs (inner Galaxy) and 230\degs (outer Galaxy) and their latitudinal value was a strong function of longitude. In addition, the FILM data was smoothed to 1\degs to improve the signal to noise. Therefore, FILM mainly probed the local ISM (at 1 kpc $b$ = 5\degs  corresponds to $\sim$90 pc, above the bulk of the \cii emission in the plane and is not likely representing high $z$--distances.)  Without   spectral resolution \cite{Nakagawa1998} were unable to assign a distance to the \cii emission and derive a spatial profile. While the GOT C+ survey is spectrally resolved, it was limited to $|b|$ $\le$  1\deg, which is only $\sim$150 pc at the distance to the Galactic center and so  probes well the distribution only up to $z$ = $\pm$150 pc.

Recently, \cite{langer2014_z} estimated the average vertical \cii $z$--scale height by ``inverting''  the BICE $b$--scan data using {\it a priori} knowledge of the radial distribution of Galactic \cii emission  at $b$=0\degs derived  from GOT C+ data and showed that the total \cii emission (from all gas components) has a broader $z$--distribution,  FWHM $\sim$172 pc,  than CO ($\sim$110 pc), but narrower than \hi ($>$230 pc).   However, since the basic vertical \cii data used in that analysis is the $b$--scan  from BICE, it does not shed any light on the $z$--distribution of individual \cii gas components, especially important for the diffuse \cii  arising from the CO-dark H$_2$ gas and the WIM, which are believed to have larger scale heights.  GOT C+ surveyed \cii sparsely in latitude  at only $b$=0\deg, $\pm$0.5\deg, and $\pm$1.0\deg, and the approaches used in \cite{Pineda2013} and \cite{langer2014_II}  only studied  some of the global properties of Galactic \cii emission as a function of Galactocentric radius.  However,  the spaxel analysis approach in this paper allows us to use the data in all latitudes $b$, combined with the $V_{lsr}$--Distance to each spaxel to derive the $z$--distributions.

To derive the $z$--distribution we resample the spaxels in 9 $z$-bins, from   $z$  = -160 pc to +160 pc,  appropriate to each spaxel's observed latitude and $V_{lsr}$-Distance. In each $z$-bin we sum the (i) intensities of  \ciis, $^{12}$CO, and  $^{13}$CO detections and (ii) intensities of the \cii gas components within each \cii spaxel as described in Section 4. Because the sampling in $b$ is heavily limited to only 5 discrete latitude values   there is no uniformity of the sampling in $z$.     For example nearly 1/3 of all spaxels populate the bin at $z$ =0 pc while the rest are distributed irregularly among the other 8 $z$-bins.  Therefore,   as discussed in Section 4.1.2, to study the relative intensity distributions of each emission  component  as a function of $z$ we need to ``normalize'' appropriately the sum of intensities in each $z$--bin.   While, in principle, each spaxel (with fixed velocity width of 3 \kms) represents a Galactic volume element appropriate to its $l$, $b$,  $V_{lsr}$ and the beam area, it varies widely from spaxel to spaxel  (especially near the tangencies) and is  uncertain. Therefore, an absolute volume normalization will be very unreliable. Instead as discussed in Section 4.1.2 we use the number of \hi  spaxels in a given $z$--bin to represent a measure of the Galactic volume   sampled  within that $z$--bin and use this for normalization.   As long as we have a sufficient number of \hi spaxels in each $z$--bin we can derive a reasonable relative intensity $z$-- distribution of CO, \cii and its gas components.

Any attempt to study the $z$--distribution requires reasonably good distances to individual spaxels.  Therefore, for this analysis, we use only the subset of the spaxels within R$_G <$ 8.5 kpc for which there is no near-- \space far--distance ambiguity (see Table~\ref{tab:spaxel_analysis}). We also include the spaxels for which the near-- \space far--distance ambiguity is resolved using \hi absorption data (see Section~\ref{sec:spaxel_distance}). In addition, we include spaxels for which the near-- \space and far--distances are within 2 kpc  of each other (near the tangencies).  In the latter case  the error in the distances used to locate the emission from a spaxel could be as large as 1.0 kpc with a corresponding error in $z$ of $\sim$8 pc for $|b|$= 1.0\deg, which we believe is small enough to be included in our analysis.  The number of spaxels used for determining the $z$--distribution are given in Table~\ref{tab:spaxel_analysis}.

\begin{figure}[!ht]
\hspace{-0.5cm}
\centering
\includegraphics[scale=0.430,angle=00]{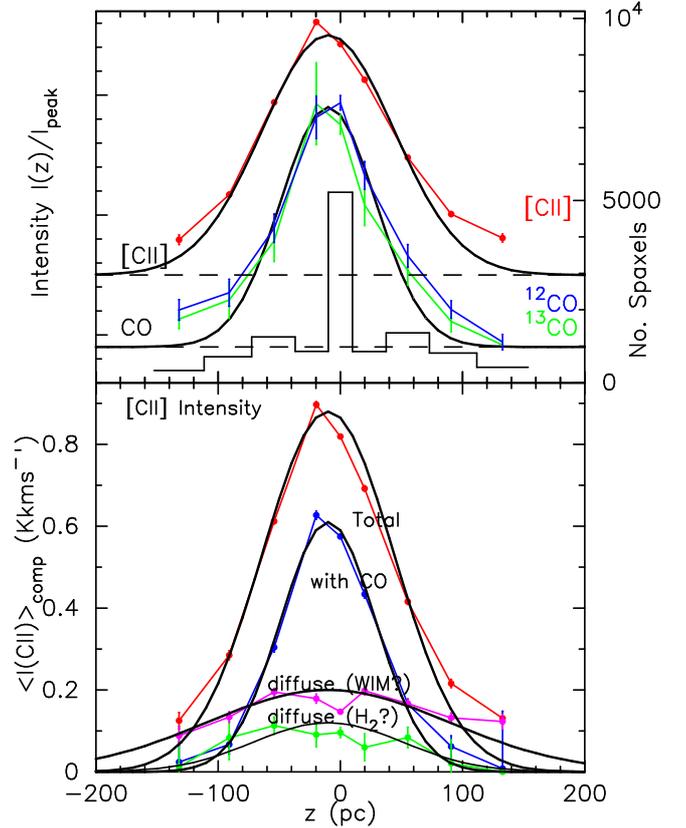}
\caption{The distributions of gas tracers as a function of $z$. (top panel) Comparison of average intensities of detections  in each $z$--bin normalized to the peak:   \cii (red) and CO (blue and green). The total number of \hi spaxels in each $z$--bin  used to represent the ``volume'' sampled for averaging are  shown by the histogram.    (lower panel) Comparison of the average intensities  of gas components of   \cii emission: total (red), with CO (blue),  (CO-faint) diffuse \h2 (green), and  diffuse (without CO) \hi and WIM (magenta).   In both panels  the thick black lines represent Gaussian fits to the data (see fits in Table~\ref{tab:z-distribution}).}
\label{fig:z-scale}
\end{figure}

\subsection{$z$-scale and FWHM distributions of \cii and CO}
\label{sec:z-scale-total}

To determine the $z$--distribution of \cii we separated the spaxels into nine  bins in $z$, as shown in the histogram at the bottom of the top panel in Figure~\ref{fig:z-scale}.  The width of  the  $z$--bins varies in order to  have as many samples in each bin as possible, while at the same time providing sufficient coverage in  $z$.  Then we summed the spaxel intensities of each tracer within each  $z$--bin.   Note that we used only the intensities in the spaxels with 3-$\sigma$ or better detections (using all the spaxels, including the ones with no signal, only  introduces more noise).  We then normalize the total intensity in each $z$--bin by dividing it by the number of \hi spaxels in it (see histogram in Figure~\ref{fig:z-scale} top panel).  As discussed above, the \hi samples provide  a good normalization for each  $z$--bin.  The error bars correspond to the 1-$\sigma$ scatter in the spaxel intensities in each $z$--bin.    Considering the availability of only a limited sampling of  spectrally resolved data in $b$, we believe that this approach,  the first ever   attempt,     provides good insight into the  $z$--distribution for \cii and the other tracers.   Following \cite{langer2014_z}, we assume a Gaussian profile to fit the emission distribution,

\begin{equation}
f(z) = f(z_c)e^{-0.5((z-z_c)/z_0)^2}
\end{equation}

 \noindent where $z_0$ is the scale height (where the full-width half-maximum, FWHM = 2(2ln2)$^{0.5}$$z_0$ = 2.35$z_0$), $z_c$  is an offset in the peak of the distribution,
 and $f(z_c)$ is the intensity at the peak in the distribution.   For the $z$-scale analysis we use all detections of $^{12}$CO and $^{13}$CO with or without \ciis, satisfying the criteria for R$_G$ and distance as for \cii and \hi spaxels. We plot  the distributions for the three gas tracers   in Figure~\ref{fig:z-scale}, for \ciis, $^{12}$CO, and $^{13}$CO, and plot the Gaussians that fit the data.  The Gauss fit parameters are given in Table~\ref{tab:z-distribution}.   As can be seen in Figure~\ref{fig:z-scale} the Gaussians provide reasonable fits to the data.  We are fairly confident of the fit to the \cii distribution because the $^{12}$CO and $^{13}$CO in Figure~\ref{fig:z-scale} and their FWHM in Table~\ref{tab:z-distribution}  are consistent with the range of 60 -- 140 pc for CO for R$_G < $R$_{\odot}$   derived from large scale Galactic CO surveys \cite[cf.][]{sanders1984,clemens1988}.   The \hi distributions (not shown here) are also consistent with the multiscale height solution summarized by \cite{dickey1990}.

  It can be seen in Figure~\ref{fig:z-scale} that the \cii relative intensity (top panel) does not go to zero at $z$=$\pm$160 pc, and thus \cii emission extends beyond this height.  This result is consistent with the BICE and FILM observations of weak \cii emission at values of $|b| >$ 1\deg, but the GOT C+ coverage is insufficient, as discussed above, to extend the distributions beyond this point. The $z$--distribution is important to calculate the \cii luminosity of the Galaxy by combining the well sampled distribution in the plane with the vertical distribution \cite[cf.][]{Pineda2014}.  Furthermore,  to calculate the luminosity of the Galaxy and study the dynamics of the different ISM components it is important to determine these individually rather than use an average value.

\begin{table}[!htbp]
\caption{$z$--distribution Gauss fit parameters:}	
\label{tab:z-distribution}	
\setlength{\tabcolsep}{0.1cm}
\begin{tabular}{lcccc}	
\hline	
Emission & peak$^\ddag$ &	FWHM & $z_c$$^\S$ &  Intensity fractions\\
 & & & & integrated in $z$\\	
  & K \kmss & (pc)  & (pc)	& (\% of \ciis$_{total})$\\													
\hline
\cii  & 0.9 & 179 & 10 & 100\\
$^{12}$CO & 6.0 &129   &10 & --- \\
$^{13}$CO &0.9 &129  &10  & ---\\
\hline	
{\bf \cii gas components:} & & & & \\
\ciis-with-CO & 0.61 &	129   &10 & 48 \\
\\
{\bf Diffuse \ciis-w/o-CO:} & & & & \\
In CO-faint \h2 & 0.12& $\sim$202 & 10 & 14\\
In \hi and WIM & 0.20 & $\sim$329& 10 & 41\\
\hline	
\end{tabular}	
\\
 $^{\S}$$z_c$  is an offset in the peak of the \cii distribution  \\
$^\ddag$peak intensity in the $z$-bins (used for normalization)\\																		
\end{table}

\subsection{$z$-scale of \cii gas components}

 Our estimate for the  $z$-scale FWHM for the total \cii of 179 pc  is in good agreement with the value of 172 pc  calculated from the inversion of the BICE $b$--scan data by \cite{langer2014_z}. However, in the analysis by \cite{langer2014_z} it was not possible to determine the scale heights of the separate gas components traced by \ciis,  because the  \cii data used in the $b$--scan is from BICE.  Whereas, in the approach adopted here, we use only the GOT C+ data, and therefore, it allows us to separate the different gas components and their distributions.    To derive their respective $z$--distributions  we follow the procedure described in Section~\ref{sec:z-scale-total}  for the total \cii emission.  We use the spaxels identified by their association with different tracers, as follows: (i) \ciis-with-CO as dense \h2 gas (see Section 4.2), (ii) bright diffuse \ciis-w/o-CO  as the diffuse CO--faint \h2 gas  (see Section 4.3.2), and (iii) low brightness diffuse \ciis-w/o-CO as diffuse \hi or WIM (see Section 4.3.2).
  To estimate the $z$--scales
 we assemble  the selected ``labeled" spaxels in $z$--bins and find their average intensity in each bin. The results are shown in the lower panel in  Figure~\ref{fig:z-scale}.  In this figure we plot the average intensities in each $z$--bin for the total \cii and in its separate gas components.  The approximate  Gaussians  to the z--distribution also are plotted in Figure~\ref{fig:z-scale} (lower panel). The Gaussian fit to the CO distribution in the top panel fits well the    z--distribution of \ciis-with-CO.     The error bars represent the 1-$\sigma$ uncertainty estimated from the spectral noise in the spaxels. The ``dips'' in the intensities of the diffuse components correspond to spaxels at, or close to, the galactic plane ($b$ = 0\degr) where  the diffuse emission could be underestimated because of the overcrowding of  the brighter \cii velocity features (mostly seen along with CO). Nevertheless, the data in Figure~\ref{fig:z-scale} brings out clearly the $z$--extent of different \cii components as indicated by  the quality of the Gaussian fits.  The Gaussian fit parameters are listed in Table~\ref{tab:z-distribution} for the separate \cii gas components.   The Gaussian fits for the diffuse components in Figure~\ref{fig:z-scale} are indicative of their broader distribution, but the exact values of the FWHM should be used with caution given the small number of samples (see the histogram in Figure~\ref{fig:z-scale}) for the $z$--bins available within the GOT C+ data and the uncertainties ($\sim$15 to 20 pc)  in the $z$--distance estimates.

 The vertical displacement  of the peak of the emissions $z_c$ $\sim$ 10 pc seem to fit all emissions and this may imply a simple displacement of the location of  the Sun with respect to the Galactic plane.  This result is consistent with the evidence of large-scale displacements of gas and young stars from a simple plane. (These displacements are significant as they are likely to be
 a consequence of 
the dynamical effects on disks in spiral galaxies \cite[e.g.,][]{Lockman1977,Matthews2008,Widrow2012}.)

\begin{table}[!htbp]																	
\caption{Overall Galactic \cii emission fractions in different gas components in the inner Galaxy R$_G$ $<$ R$_{\odot}$, integrated over the inner Galactic disk, including the $z$-scales }
\label{tab:galactic_fractions}												
\begin{tabular}{ll}																	
\hline																	
 ISM  Gas component & \cii  intensity  fraction$^1$\\
\hline	
$^2$\cii fraction in dense \h2  & 48\% \\
$^2$\cii fraction in (CO-faint) diffuse \h2  & 14\% \\
 $^3$\cii fraction in diffuse \hi and  WIM & 41\% \\
\hspace{0.8cm} \cii fraction in diffuse \hi & 10.5\% \\
\hspace{0.8cm} \cii fraction in WIM  & 30.5\% \\
\hline
\end{tabular}
\\
$^1$Includes only \cii intensities above GOT C+ detection limit\\		
$^2$Ignoring the emission from \hi layers in the \h2 clouds.\\	
$^3$Does not include emissions below GOT C+ detection limit  																	
\end{table}	

The  \cii sources associated with CO, have the narrowest distribution, FWHM = 129 pc, and, as expected, it agrees well with our value for CO emission because these are physically associated with each other, representing an identical population of CO spaxels with \cii emission.  This component makes the largest  contribution to the total \cii emission  (at peak $\sim$60\%)  but decreases sharply with $z$.   In contrast the diffuse \hi or WIM component has the largest extent (FWHM $\sim$330 pc),   as expected for \cii emission from the WIM or diffuse \hi and may be consistent with the weak emission observed by FILM at higher latitudes \cite[cf.][]{Makiuti2002}.   The scale height   $\sim$ 400 pc for  the emission measure (EM) derived from the Wisconsin H$\alpha$ Mapper (WHAM) data \cite[cf.][]{gaensler2008} also suggests  a larger $z$--distribution may be observed for the WIM component  of the \cii emission.    The \cii fraction in this   diffuse component is $\sim$22\% near the plane of the disk and decreases slowly with $z$. Therefore, the  intensity integrated over all $z$  will be much larger   and it will make a significant contribution to the Galactic \cii luminosity. The (CO-faint) diffuse \h2 gas  component has an intermediate $z$--distribution  with a FWHM $\sim$202 pc and   its peak is $\sim$13\% of the total.

 The (CO-faint) diffuse \h2 emission does  not fit well with a Gaussian, instead   its \cii fraction stays nearly constant at a level of  $\sim$12\%  from $z$ $\sim$ -100 pc to $\sim$ +60 pc and beyond that decreases faster.  This result in particular is subject to the uncertainties   due to having the fewest samplings  of  $z$--bins and is possibly underestimated because of overcrowding of bright \cii features near the plane.  Despite the smaller number of samples,  we   conclude with some confidence that  the (CO-faint) diffuse \h2 gas  traced by \cii  is significantly present from $z$ $\sim$ -140 pc to $\sim$ +100 pc.  In the case of the diffuse \h2 component it is possible that  the flatness (or deficiency) near the plane, as indicated by data points in Figure~\ref{fig:z-scale} for the $z$--distribution, is real.  Such a situation can be interpreted as a natural consequence of the time-scales for cloud evolution in and out of the Galactic plane.  Thus our results  in Figure~\ref{fig:z-scale} is consistent with this difference and would imply that evolution  from \hi clouds to diffuse \h2 clouds and finally to CO molecular clouds is more rapid in the plane than at higher distances above the plane.    Finally, we note that the distribution of the diffuse \hi and/or WIM emission, having the largest FWHM, likely dominates the \cii emission above the plane and is probably responsible to the high latitude emission seen by BICE and FILM.

We use the intensity for each gas component in Figure~\ref{fig:z-scale} and  their respective Gauss fits in Table~\ref{tab:z-distribution} to estimate the overall \cii intensities and gas fractions integrated  over the inner Galactic disk both radially and vertically.    This  calculation  is possible because the intensities at $z$ = $z_c$  represent an   average over  the entire inner Galaxy R$_G$ $<$ R$_\odot$.    We now know the form and extent of this emission in the vertical direction $z$ from   Equation (2).  Then, by using the intensities and their FWHM listed in Table~\ref{tab:z-distribution} we obtain the \cii gas fractions averaged  over the entire  inner Galaxy including their  $z$--scales.    These estimates are summarized in  Table~\ref{tab:galactic_fractions}. Although we do not fit $z$--distributions separately for the diffuse \hi and WIM, in Table~\ref{tab:galactic_fractions},  we list their  \cii gas fractions averaged  over the entire  inner Galaxy, as estimated   using the results in Section 4.3.3 for the  ratio \ciis$_{\rm WIM}$ to \ciis$_{\rm {H I}}$ in the diffuse ISM ($\sim$3), and assuming they have the same $z$-scales.   However, our estimates are only relative to the averaged \cii intensities and are not in absolute units representing a physical volume (as in pc$^3$). When  given in absolute volume units this data can  be readily compared with the brightness and luminosity in other galaxies. Nonetheless, these relative intensities (which includes the differences in their $z$-scale heights) as  fractions of the total \cii are still   valuable for understanding  the \cii  emission in the Galaxy as well as in external galaxies. If we have an estimate of the total \cii luminosity of the Galaxy (e.g., \cite{Pineda2014} derived 10.1$\times$10$^{40}$ ergs s$^{-1}$) we can calculate the luminosity in each gas tracer  for comparison to external galaxies by scaling the luminosity using the \cii gas fractions in Table~\ref{tab:galactic_fractions}.


\section{Summary}
In this paper we show evidence of the  prevalence of a diffuse \cii emission component detectable in the HIFI position-velocity maps and in the line-wings of individual spectra. To provide a global perspective of distribution and origin of the gas components of \cii emission in the Galaxy we  made a comprehensive (3--\kmss  wide) spaxel by spaxel profile analysis of all GOT C+ survey and ancillary spectral line data toward the inner Galaxy (R$_G$ $< $ R$_{\odot}$),  and characterized all the \hi  velocity features with and without \cii detections.
Our approach differs from previous analysis of the GOT C+ data base in that here we characterize the emission of \cii and auxiliary CO, \cis, and \hi in each spaxel  along with kinematic distances derived from their $V_{lsr}$.    We derive their 2-D distributions in the plane of the Galactic disk and in $z$. We find that the \cii emission is correlated with the spiral arms in the Galaxy. We then analyze spaxel by spaxel  the correlation of \cii spaxel intensity to that of other gas tracers to identify the likely source of the diffuse \cii emission.

   The spaxels with \cii detections were further divided into \ciis-with-CO and \ciis-w/o-CO.   With this analysis we separated the diffuse \cii component from that in the denser molecular \h2 gas and estimated the \cii fraction in them (defined as the ratio of  intensity in each gas component to the total \ciis). The diffuse \cii emission was further classified  by its possible identification with   (i) \cii in the diffuse molecular \h2 clouds (CO-faint) with a fraction $\sim$13\% and (ii) with \cii in diffuse \hi clouds and the warm ionized medium (WIM) with a fraction  $\sim$ 26\%. We estimate that diffuse \hi gas contributes $\sim$  6\% to diffuse \ciis, thus the remainder  ($\geq$ 21\%)  comes from  the  WIM. The estimates of  \cii  gas fractions derived using the spaxel intensities alone are given in Figure~\ref{fig:summary} and the estimates of the  \cii gas fractions that include their z--scales are listed in Table~\ref{tab:galactic_fractions}.

  Our results  are broadly consistent with those of  \cite{Pineda2013} who calculated the \cii emissivity in the plane for $b$=0\deg, except for the WIM.   However, our determination for the WIM contribution comes directly from the data, whereas \cite{Pineda2013} calculated only 4\% for the WIM  emissivity for $b$=0\deg, using an electron density model which  underestimates the WIM contribution to \cii emission as discussed in Section 4.3.3. In the diffuse ISM we estimate the ratio \ciis$_{\rm WIM}$ to \ciis$_{\rm {H I}}$ $\sim$3, which makes the WIM the dominant source for diffuse \cii observed with GOT C+. The large fraction of WIM contribution to \cii is further corroborated by the extended $z$--scale of the diffuse \cii component.

We estimate about 62\% of the total \cii  intensity in the inner Galaxy is produced in \h2 gas. In other words, this fraction of \cii intensity {\it traces} \h2 molecular gas clouds.   Furthermore,  it is possible, under extremely low  \c+  excitation conditions such as at low kinetic temperatures ($<$ 30K) and/or densities, in their C$^+$--\h2  layers, that the \cii emission is not detected.   In the GOT C+ survey not all CO detections have associated  \cii emission and thus a significant fraction of the Galactic \h2 gas is missed by \cii as a \h2 gas tracer.     Using CO as a \h2 gas tracer,  we  can  estimate  the fraction  of \h2 molecular clouds missed by \cii from the  spaxels with CO and those with CO but no \cii   detections  (see Table~\ref{tab:spaxel_analysis}). A rough estimate of the fraction of the number of \h2 clouds missed by \cii as a tracer is  $\sim$ 38\%  suggesting a significantly large number of CO clouds are too cold to detect \cii   at the sensitivity of the GOT C+ survey.

   We show that a comparison of CO with and without associated \cii separates the warm and cold \h2 in the Galaxy. From their radial distributions the warm-\h2 gas fraction   ($\sim$ 0.6 to 0.7)  peaking in the range 4 kpc $<$ R$_G$ $<$ 7 kpc  is consistent with the peaks in the star formation rate and star formation efficiency,  and provides an important observational constraint on Galactic star formation.

  One of the limitations  in utilizing the GOT C+ data set for analyzing the $z$--distribution, is  that it is  a sparse survey covering longitudes from 0\degs to 360\deg,  and  only   a few latitudes $|b|$ = 0\deg, $\pm$0.5\deg, and $\pm$1.0\deg.  A major difficulty in analyzing such a sparse data set is that not all of the Galactic volume is sampled equally, thus a quantitative analysis requires being able to normalize the sampling volumes throughout the Galaxy.  We do not attempt  to calculate a geometrical volume,  which would be difficult to do, instead we have developed a normalization   procedure or technique  that   uses the spaxels with \hi emission to sample and measure an effective volume in the Galaxy.
With this approach we are able, for the first time,  to separate out the $z$--distribution of all ISM components traced by \ciis:  (i) dense molecular \h2 clouds, associated with CO, having the narrowest distribution, FWHM = 129 pc  similar to that of CO; (ii) the CO-faint diffuse molecular  clouds being slightly broader, FWHM $\sim$ 202 pc; and, (iii)  the broadest distribution is found for  the diffuse \cii arising from the atomic \hi and$/$or ionized gas (WIM), with a FWHM $\sim$330 pc. All the fits and offsets are given in Table~\ref{tab:z-distribution} along with a summary of the fraction of Galactic \cii emission in these gas components.   The $z$--distributions are subject to the uncertainties in resolving the near-- \space far--distance ambiguity in the  spaxel kinematical distances.   Though these results are  subject to the ambiguities in distances, the use of a large ensemble of spaxels in our statistical analysis  reduces  their effect on the result.

  Calculating the Galactic \cii gas fractions without the inclusion of $z$--scales tends to overestimate the fraction in dense molecular gas due to the dominance of molecular clouds in the plane ($b$=0\deg) while underestimating the diffuse \cii components. Thus our results on the \cii gas fractions, including their $z$-scales,  provide  a more robust estimate of the Galactic luminosity.  Though these estimates of  gas fraction are in  relative intensities (which includes the differences in their $z$--scale heights)  they are still   valuable for understanding  the \cii  emission in the Galaxy as well as in external galaxies when used with the total \cii luminosity \cite[cf.][]{Pineda2014}.  The GOT C+ data,   publicly available in the {\it Herschel} archives  provides  a  useful resource for any future work on \c+ emission.

\begin{acknowledgements}
 We thank the referee Dr. Jay Lockman for a careful reading of the manuscript and for constructive comments and suggestions. We thank the staffs of the ESA {\it Herschel} Science Centre and NASA {\it Herschel} Science Center, and the HIFI, Instrument Control Centre (ICC)  for their help with the data reduction routines.
This work was performed at the Jet Propulsion Laboratory, California Institute of Technology, under contract with the National Aeronautics and Space Administration.

\end{acknowledgements}

\bibliographystyle{aa}
\bibliography{aa_CII_diffuse_refs}
\end{document}